\documentclass[twoside,12pt]{Latex/Classes/PhDthesisPSnPDF}

\newcounter{bookfootnote}

\ifpdf  
    \pdfcatalog { /PageMode (/UseOutlines)
                  /OpenAction (fitbh)  }
\fi

\title{CMB Tensor Anisotropies\\ in $f(R)$ Gravity}

\ifpdf
  \author{\href{mailto:hassan.bourhrous@uct.ac.za}{Hassan Bourhrous}}
  \collegeordept{\href{http://www.mth.uct.ac.za/}{Department of Mathematics}}
  \university{\href{http://www.uct.ac.za}{University of Cape Town}}

  \crest{\includegraphics[width=4.5cm]{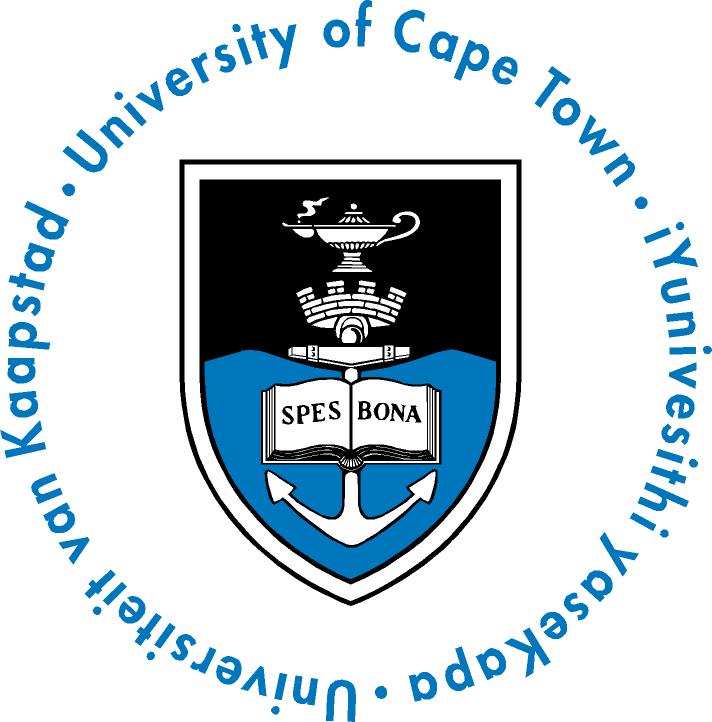}}

\else
  \author{Hassan Bourhrous}
  \collegeordept{Department of Mathematics}
  \university{University of Cape Town}
  \crest{\includegraphics[width=4cm]{uct_logo.png}}
\fi

\degree{MSc in Theoretical Cosmology}
\degreedate{December 2012}

\newcommand{\fR}{$f(R)$ }
\newcommand{\Rn}{$R^n$ }
\newcommand{\ph}{\phantom}
\def \vec #1{\textbf{\em  #1}}
\newcommand{\Cl}{C_l} 
\newcommand{\Pl}{P_l} 

\newcommand{\la}{\langle}
\newcommand{\ra}{\rangle}
\newcommand{\calR}{\mathcal{R}}

\newcommand{\cinc}{\stepcounter{bookfootnote}} 
\newcommand{\countval}{\value{bookfootnote}}   

\def\edth{\;\raise1.0pt\hbox{$'$}\hskip-6pt\partial\;}
\def\baredth{\;\overline{\raise1.0pt\hbox{$'$}\hskip-6pt
\partial}\;}
\def\bi#1{\hbox{\boldmath{$#1$}}}
\def\gsim{\raise2.90pt\hbox{$\scriptstyle
>$} \hspace{-6.4pt}
\lower.5pt\hbox{$\scriptscriptstyle
\sim$}\; }
\def\lsim{\raise2.90pt\hbox{$\scriptstyle
<$} \hspace{-6pt}\lower.5pt\hbox{$\scriptscriptstyle\sim$}\; }

\newcommand{\p}{\partial}

\def\l{\left}
\def\r{\right}

\def\mat#1  { \begin{matrix}#1\end{matrix} }
\def\pmat#1 { \begin{pmatrix}#1\end{pmatrix} }
\def\cas#1  { \begin{cases}#1\end{cases} }

\def \D {\tilde{\nabla}}
\def\3nab{\tilde{\nabla}}
\def\lgl{\langle}
\def\rgl{\rangle}
\def\nn{\nonumber}

\newcommand{\curl}{\,\mbox{curl}\,}

\numberwithin{equation}{chapter}

\begin{document}


\renewcommand\baselinestretch{1.2}
\baselineskip=18pt plus1pt


\maketitle  


\newpage
\vspace{10mm}
Supervisor: Prof. Peter K. S. Dunsby

\vspace{10mm}
Reviewer 1: 

\vspace{10mm}
Reviewer 2: 

\vspace{10mm}
External Examiner:

\vspace{20mm}
Submission date: December 19, 2012

\vspace{20mm}
\hspace{70mm}Committee Approval:


\frontmatter
\setcounter{page}{5}

\begin{dedication} 

To my family

\newpage

\flushleft
\phantom{Tajdidt n'Aari}
\vspace{2cm}
\hspace{2.5cm} {\bf \Huge Tajdidt n'Aari}\\
\hspace{2.5cm}\includegraphics{tajdidt}

\vspace{0.2cm}
\hspace{2cm} Mots doux de mains port\'ees de l\`evres en brises,\\
\hspace{2cm} Mots que tirent ses doigts secs et fins\\
\hspace{2cm} Mots malax\'es comme h\'esitants\\
\hspace{2cm} Mots lib\'er\'es soudain, offerts d'un geste tendre\\
\hspace{2cm} Mots port\'es, offerts au monde, \`a l'autre, au vent\\
\vspace{0.2cm}
\hspace{2cm} Toute petite grande dame,\\
\hspace{2cm} Ecoute les mains de Fettouma,\\
\hspace{2cm} Elles ont la m\'emoire des b\^atons noueux, la m\'emoire\\
\hspace{2cm} des pierres et des lessives glac\'ees,\\
\hspace{2cm} Elles ont la force sure du monde qu'elles embrassent\\
\hspace{2cm} D'un monde r\'eduit \`a la terre de ses pas\dots et immense\\
\vspace{0.2cm}
\hspace{2cm} pourtant \ldots\\

\hspace{9cm} G. Drumont\\
\hspace{9cm} Ao\^ut 2010\\

\end{dedication}


\begin{acknowledgements}      

I would like to thank all those who contributed to this work in one way or another, particularly:\\
Prof. Peter Dunsby for an excellent supervision.\\
Dr. \'Alvaro~de~la~Cruz-Dombriz.\\
Mr. Mohamed Abdelwahab.\\
Prof. Hassane Darhmaoui.\\
Dr. Garry Angus.\\
Dr. Rituparno Goswami.\\
Ms. Anne-Marie Nzioki.\\
Dr. Julien Larena.\\
Mr. Amare Abebe.\\
Dr. Bernard Leong.\\
Dr. Rockhee Sung.\\

HB acknowledges NASSP/NRF funding.
\end{acknowledgements}




\begin{abstracts}        
The cosmic microwave background (CMB) carries information from the last scattering surface that puts constraints on the multitude of proposed cosmological models and the gravitation theories they are based on. One class of such theories is $f(R)$ gravity, which has become an interesting endeavour to correct for the degeneracies of the concordance model.\\
We presents a description of CMB anisotropies generated by tensor perturbations in $f(R)$ theories of gravity. The temperature and the $E$-mode polarisation power spectra in the special case of $f(R)=R^n$ are computed using a modified version of CAMB package.
\end{abstracts}




\begin{declaration}        
This dissertation on \emph{CMB tensor anisotropies in $f(R)$ theories of gravity} is submitted for the degree of master of science in theoretical cosmology. I hereby declare that this work is my own and was completed without prohibited assistance of third parties and without making use of aids other than those specified. Extracts from other sources have been identified as such and clearly referenced. 

The work was carried out during the period from June 2010 through July 2012 under the supervision of Prof. Peter Dunsby at the department of Mathematics, University of Cape Town.

\hspace{7.6cm} Hassan Bourhrous

\vspace{-2mm}
\hspace{7.6cm} Azrou --  Morocco

\vspace{-2mm}
\hspace{7.6cm} December 19, 2012
\end{declaration}



\begin{tiny} 
\printnomenclature[1.5cm] 
\label{nom} 
\end{tiny} 
\markboth{{\nomname}}{{\nomname}}

\chapter*{Convention and Notation}
\section*{Convention}
\begin{small}
\begin{tabbing}
\hspace{5cm}\= \\
Metric signature: \> $(-,+,+,+)$\\
Speed of light: \> $c=1$ \\
$\kappa$: \>  $\frac{8 \pi G}{c^2} = 1$
\end{tabbing}
\end{small}

\section*{Notation}
\begin{small}
\begin{tabbing}
\hspace{2cm}\= \\

$abc\cdots$:  \> Spacetime indices $\in \{0,1,2,3\}$\\
$ijk\cdots$:  \> Spatial indices $\in \{1,2,3\}$\\
$(ab\cdots)$:  \> Symmetry over enclosed indices $ab\cdots$\\
$[ab\cdots]$:  \> Anti-symmetry over enclosed indices $ab\cdots$\\
$\langle ab \rangle$:	\> Orthogonal projection of the symmetric trace free part over indices $a$ and $b$\\
$\partial_a$ or ,$a$ :  \> Partial derivative $\l(\equiv \frac{\p}{\p x^a}\r)$\\
$\nabla_a$ or ;$a$ :  \> Covariant derivative with respect to $g_{ab}$\\
$\bar{\nabla}_a$:  \> Covariant derivative with respect to the affine connection\\
$\tilde{\nabla}_a$:  \> Projected covariant derivative on 3-space\\
$\eta_{abcd}$: \> Totally anti-symmetric tensor on spacetime\\
$\epsilon_{abc}$: \> Projected totally anti-symmetric tensor ($\equiv \eta_{abcd}u^d$)\\
$a$: \> Scale factor\\
$t$: \> Cosmic time\\
$\eta$: \> Conformal time ($\text d \eta \equiv \frac{\text d t}{a}$)\\
$\tau$: \> Proper time\\
$x_a$: \> Position\\
$u_a$: \> 4-velocity ($\equiv \frac{\text d x_a}{\text d \tau}$)\\
$A_a$: \> 4-acceleration\\
$g_{ab}$: \> Metric tensor\\
$g$: \> Determinant of the metric tensor\\
$\gamma_{ij}$: \> 3-metric tensor\\
$h_{ab}$: \> Space projection tensor ($\equiv g_{ab} + u_a u_b$)\\
$U_{ab}$: \> Time projection tensor ($\equiv - u_a u_b$)\\
$\{^c_{ab}\}$:  \> Levi--Civita connection ($\equiv \frac{1}{2} g^{cd} (g_{ad,b} + g_{bd,a} - g_{ab,d})$)\\
$\Gamma^{c}_{\phantom{c}ab}$:  \> Affine connection\\
$T_{ab}^{\phantom{ab}c}$:  \> Torsion tensor ($\equiv 2 \Gamma^c_{\phantom c [ab]}$)\\
$Q_{abc}$:  \> Non-metricity tensor ($\equiv - \bar \nabla_a g_{bc}$) \\
$R_{abcd}$: \> Riemann tensor constructed from the metric tensor (\S \ref{manifolds})\\
$R_{ab}$: \> Ricci tensor ($\equiv g^{cd} R_{cadb}$)\\
$R$: \> Ricci scalar ($\equiv g^{ab}R_{ab}$)\\
$\calR_{abcd}$: \> Riemann tensor constructed from the affine connection\\
$\calR_{ab}$: \> $\equiv g^{cd} \calR_{cadb}$\\
$\calR$: \> $\equiv g^{ab} \calR_{ab}$\\
$C_{abcd}$: \> Weyl Tensor\\
$E_{ab}$: \> Electric Weyl tensor ($\equiv u^c u^d C_{acbd}$)\\
$H_{ab}$: \> Magnetic Weyl tensor ($\equiv \frac{1}{2} \, \epsilon_{acd} \, u^e C_{be}{}^{cd}$)\\
$\mathcal{L}_m$: \> Matter Lagrangian \\
$S_m$: \> Matter action \\
$\Psi_m$: \> Matter field \\
$T_{ab}$: \> Energy-Momentum tensor $\left(\equiv -\frac{2}{\sqrt{-g}}\frac{\delta \mathcal{L}_m}{\delta g^{ab}}\right)$\\
$\Delta^c_{\phantom cab}$: \> Hypermomentum $\left(\equiv -\frac{2}{\sqrt{-g}}\frac{\delta \mathcal{L}_m}{\delta \Gamma^c_{\phantom cab}}\right)$\\ 
$\Theta$: \> Expansion parameter ($\equiv \D^a u_a$)\\
$\rho$: \> Energy density\\
$p$: \> Energy (isotropic) pressure\\
$w$: \> Equation of state parameter $\left(\equiv \frac{p}{\rho} \right)$\\
$q_a$: \> Energy flux\\
$\sigma_{ab}$: \> Shear ($\equiv \D_{\la a} u_{b \ra}$)\\
$\omega_{ab}$: \> Vorticity ($\equiv \D_{[a}u_{b]}$)\\
$\omega_a$: \> Vorticity vector ($\equiv \epsilon_{abc} \omega^{bc} /2 = \curl u_a /2$)\\
$\pi_{ab}$: \> Anisotropic stress\\
$c_s$: \> Sound speed $\left( c_s^2 \equiv \frac{\dot p}{\dot \rho} \right)$\\
$G$: \> Newton's gravitational constant
\end{tabbing}
\end{small}

\markboth{{Convention and Notation}}{{Convention and Notation}}




\setcounter{secnumdepth}{3} 
\setcounter{tocdepth}{1}    
\tableofcontents            
\markboth{{Contents}}{{Contents}} 


%

%
%


\mainmatter

\renewcommand{\chaptername}{} 




\chapter{Introduction}
\label{introduction}
\ifpdf
    \graphicspath{{1_introduction/figures/PNG/}{1_introduction/figures/PDF/}{1_introduction/figures/}}
\else
    \graphicspath{{1_introduction/figures/EPS/}{1_introduction/figures/}}
\fi



The early universe was the play ground of the dynamics that led to the large scale structure observed today. It naturally follows that understanding the Universe requires deep digging into its history. To do such a thing, one needs a cosmological model (thus a theory of gravity) that meets the observations. Currently, the cosmic microwave background (CMB) is arguably the best constraint on big bang cosmological models, to which it was the greatest proof. Over the years, the measurements of the CMB have improved and with them our understanding of the origin and the evolution of the Universe \nocite{Dunsby:1997xy,2001ctap.conf..177E,Challinor:1999xz,Dodelson} \cite{Dunsby:1997xy}--\cite{Dodelson}.

The CMB radiation comes to us from a region in spacetime dubbed \emph{the surface of the last scattering} (SLS). Located at redshift $z \approx 1100$, it is closely associated with the time when the temperature of the Universe became low enough to allow the capture of electrons by hydrogen nuclei in what is known as \emph{recombination}. The physical importance of the CMB is in the sense that the duration of recombination is negligible compared to the total age of the Universe. That is to say that the surface of the last scattering is actually a shell of negligible thickness $\Delta z \approx 0.067$ of the mean redshift \cite{Dunsby:1997xy,2001ctap.conf..177E}. Before recombination, radiation was strongly coupled to matter via Thompson scattering. The Universe, therefore, was very homogeneous. After recombination, the mean free path of the photons increased significantly as there were hardly any electrons to scatter them. They have been travelling freely since then while continuously loosing energy due to the expansion of the Universe. Rarely interacting, the CMB pristinely encodes a wealth of information about the early universe \cite{Dodelson,Zaldarriaga:2003bb}.

A few authors, e.g. \cite{PhysRev.75.1089}, theorised about the CMB as a direct consequence of the big bang and its temperature. Many experiments were devised in effort to detect the relic radiation. The actual discovery came rather serendipitously in 1965 by Penzias and Wilson \cite{Penzias:1965wn} when their antenna picked up an excess temperature they could not account for. The meaning of this was immediately recognized by Dicke \emph{et al.} (1965) \cite{Dicke:1965} (see \cite{Wilkinson:1999} for history). 

Since then, there have been many more measurements of the temperature of the CMB (mostly in narrow bands) that were all crowned by the remarkably precise measurement of the CMB spectrum by the Far-InfraRed Absolute Spectrophotometer (FIRAS) instrument on board of the COsmic Background Explorer (COBE) satellite; making it the most perfect blackbody spectrum ever observed with temperature amplitude of $T_0 = 2.725 \pm 0.020$ K \cite{Smoot:1998jt}. Although FIRAS only observed the blackbody peak temperature and spectral distortions were possible, the deviations from the blackbody spectrum are expected to be small \cite{Gawiser:2000az}. It is hence possible to determine other spectral characteristics that depend only temperature such as the spectral intensity and the corresponding spectral brightness \cite[p.~8]{Naselsky}.

COBE's biggest triumph is the detection of the primordial anisotropies in the CMB, predicted two decades earlier by \cite{Sunyaev:1970eu} and \cite{Peebles:1970ag} to be in the order of $10^{-5} \lesssim \Delta T / T \lesssim 10^{-4}$. The Differential Microwave Radiometer (DMR) on  board confirmed the order of the fluctuations to be $10^{-5}$\cite{Smoot:1998jt}. It turns out that the CMB is not smooth after all and temperature varies slightly around the monopole value of $T_0 = 2.725$ K. The concept of CMB anisotropies is crucial because they represent the primordial fluctuations, the matter local densities, and the physical processes that resulted in the present structure of the Universe \cite{Hu:2002aa,Zaldarriaga:2003bb}.

The CMB blackbody spectrum undergoes further distortions after decoupling due to other effects such as the Sunyaev-Zel'dovich (SZ), the Ostriker-Vishniac (OV), the Integrated Sachs-Wolfe (ISW), and gravitational lensing. The reader is referred to \cite{Refregier:1999ri} for a review. Throughout this text, the term ``anisotropies'' refers to the primordial/primary temperature anisotropies that occurred at the SLS. Secondary anisotropies are out of the scope of this text.

The information from temperature anisotropies is richly complemented by another property of the CMB, anisotropy polarisation (See \cite{Hu:1997hv} for a good introduction). It arises from Thompson scattering accompanied with local quadrupole anisotropies at the SLS \cite{Gawiser:2000az}. Linear polarisation of the CMB, characterized by the Stokes parameters $Q$ and $U$, can be decomposed into a curl free and a divergence free components called the $E$-modes and the $B$-modes respectively, in analogy with electric and magnetic vectors in electromagnetism ($\vec E$ and $\vec B$). The importance of this decomposition lies in the fact that density (scalar) perturbations at the SLS produce only $E$-mode polarisations while the gravitational (tensor) perturbations produce both $E$ and $B$ modes in equal magnitudes. Detection of the primordial $B$-mode polarisation would directly imply the existence of gravitational waves (GW). The $E$-mode signal contributes about $5 \mu$K to the CMB temperature while the $B$-mode contribution is theorised to be $\lesssim 0.2 \mu$K \cite{Challinor:2006yh}. Measuring the $B$-modes is difficult not only because of their low magnitude but also because of the foreground contamination due to weak gravitational lensing that can redistribute some of the power of the $E$-modes into $B$-modes even in the absence of gravitational waves \cite{Hu:2002aa,Zaldarriaga:2003bb}.

Temperature, anisotropies, and polarisation are the main measurable attributes of the CMB. Other areas that can be investigated through the CMB include weak gravitational lensing, non-linear ISW, neutrino masses, cosmic strings, and primordial magnetic fields \cite{Challinor:2006yh}. CMB anisotropies are the main focus of this work. They will be discussed further below; first in general and later in \fR theories of gravity. 

Cosmological models have to be predictive in order to be testable provided the necessary technology exists. Specifically, the theories of gravity they are based on must be able to explain gravitational phenomena and cosmological dynamics \cite{Capozziello:2007ec}. Although the latter may as well be classified under the former, it makes a subtle distinction between astrophysical and cosmological scales. At large scales, tests of the theory of gravity are actually tests of the cosmological model \cite{Uzan:2010ri}. In this sense, Einstein's theory of general relativity (GR) \cite{Einstein:1915by,Einstein:1917ce}  has been very successful at the level of the solar system. The validity of GR at cosmological scales is still yet to be confirmed or refuted.

Nowadays, the most widely accepted model of cosmology is $\Lambda$CDM, a Friedmann-Lema\^itre-Robertson-Walker (FLRW) background supplemented with small perturbations and a cosmological constant term \cite{Clarkson:2010uz}. According to $\Lambda$CDM, also known as the \emph{Concordance Model}, our Universe is composed mainly of 24\% cold dark matter (CDM), and 72\% dark energy (DE) manifesting as the cosmological constant $\Lambda$ in the Einstein field equations. The remaining 4\% is the familiar baryonic matter \cite{Dunsby:2010ts,Peebles:1998yv}. 

The idea of a scalar field ($\Lambda$) driven acceleration of the expansion of the Universe is quite uncomfortable because of the completely unknown nature of dark energy. Under these circumstances, the possibility of the geometrical origin of the acceleration is an attractive one. Some theories, based on this line of thought, strive to give rise to cosmologies that evolve naturally towards late acceleration of the expansion \cite{Ananda:2007xh}. 

Departures from $\Lambda$CDM present us, in essence, with two trends. The first is abandoning the cosmological principle assumption, which asserts that we do not live in a special place nor time (See \S \ref{cosmo_princip}). Although isotropy is verifiable, homogeneity cannot be directly observed with the present means because that would require simultaneous measurements at a minimum of two other points in space separated by cosmological distances \cite{Clarkson:2010uz}. Inhomogeneous cosmologies \cite{Ellis:2011hk,Hellaby:2009vz} become the correct theory platforms within this paradigm. The second is modifying gravity while maintaining a FLRW universe \cite{Dunsby:2010ts}. This is the paradigm of \emph{extended theories of gravity} (ETGs).

Most ETGs involve higher order corrections to the Einstein-Hilbert action in such way that the classical GR may be recovered in the weak field limit. Higher order geometrical invariants such as $R^n , R_{ab}R^{ab} , R_{abcd}R^{abcd}, R \Box^k R$ \cinc \footnote[\countval]{The Ricci scalar $R$, the Ricci tensor $R_{ab}$, and the Riemann tensor $R_{abcd}$ are introduced in the following chapter.}, and minimal/non-minimal coupling terms between scalar fields and the dynamics such as $R \phi^2$, are considered in the gravitational Lagrangian. The potential of ETGs to naturally overcome the degeneracies of the concordance model such as inflation and DE; yet being fully capable of fitting observations makes them promising theories holding viable models rather than a mathematical curiosity \cite{Capozziello:2009nq}.

One subset of ETGs are the \fR theories of gravity obtained, as the name suggests, by making the gravitational Lagrangian an arbitrary function of the Ricci scalar $R$. It turns out they present a nice middle ground between simplicity and generality which makes \fR based cosmologies good models to gain insight into modified gravity. These theories will be elaborated on, subsequently, as part of the way towards establishing a theoretical framework for computing the CMB tensor anisotropies in \fR gravity, the main point of this work. 

Not only the CMB is a good means to constrain and compare cosmological models, the topic is also interesting in its own right, considering the physics involved and the rich phenomenology of the experimental data. We present a description of CMB anisotropies generated by tensor perturbations in \fR theories of gravity. Then we compute the power spectra of the observables $TT$ and $EE$ in the special case of $f(R)=R^n$ using a modified version of CAMB package \cite{Lewis:2000PhD}.

The outline of this text is as follows: Einstein's theory of general relativity and the standard cosmological model are presented in Chapter \ref{lcdm}. Then, CMB perturbations are described in Chapter \ref{cmb}. This is done following the $1+3$ covariant approach summarized therein. Chapter \ref{fR} is an overview of \fR gravity and its dynamics. Connecting the previous two chapters, CMB anisotropies in \fR gravity are established in Chapter \ref{anis_fR} together with the special case of $f(R) = R^n$. Finally, results of the performed simulations, the CMB power spectra, and the discussion follow in Chapter \ref{discussion}.


\chapter{The Concordance Model} 
\label{lcdm}


\ifpdf
    \graphicspath{{2/figures/PNG/}{2/figures/PDF/}{2/figures/}}
\else
    \graphicspath{{2/figures/EPS/}{2/figures/}}
\fi


\section{An Overview of Manifolds}
\label{manifolds}
In general, an $n$-dimensional manifold is a locally Euclidean topological space. That is to say that the neighbourhood of every point is topologically similar to an open unit ball in $\mathbb R^n$ \cite{todd_manifold}. More precisely, a manifold is any set that can be continuously parametrised. The number of independent parameters required to specify any point is the dimension of the manifold and the parameters themselves are the coordinates of the point \cite{Hobson}.

The metric tensor is defined on a manifold as:
\begin{equation}
g_{ab} \equiv \vec e_a \cdotp \vec e_b \ ,
\label{metric}
\end{equation}
where the vectors $\vec e_a$ are a local basis of the manifold.

The covariant derivative on the manifold:
\begin{equation}
T_{abc\cdots;z} = T_{abc\cdots,z} - \Gamma^k_{\phantom kaz} T_{kbc\cdots}  - \Gamma^k_{\phantom kbz} T_{akc\cdots} - \cdots  \ ,
\end{equation}
obviously requires the \emph{affine connection} \cite{Clifton:2011jh}:
\begin{equation}
\Gamma^a_{\phantom abc} \equiv  \{^a_{bc}\} + K^a_{\phantom abc} + L^a_{\phantom abc} \ ,
\label{Gamma} 
\end{equation}
where $\{^a_{bc}\}$ is the Levi-Civita connection defined in terms of the metric tensor by:
\begin{equation}
\{^{a}_{bc}\} \equiv \frac{1}{2} g^{ad} (g_{bd,c} + g_{cd,b} - g_{bc,d}) \ ,
\label{LCC} 
\end{equation}
$K^a_{\phantom a b c}$ is the contorsion tensor defined in terms of the metric tensor and the anti-symmetric parts of the connection as:
\begin{equation}
K^a_{\phantom a b c} \equiv \Gamma^a_{\phantom a [b c]} - \Gamma^d_{\phantom d [b e]} g^{ae} g_{cd} - \Gamma^d_{\phantom d [c e]} g^{ae} g_{bd} \ ,
\label{contorsion} 
\end{equation}
or in terms of the torsion tensor $T_{ab}^{\phantom{ab}c} \equiv 2 \Gamma^c_{\phantom c [ab]}$ \cite{ortin2007} as:
\begin{equation}
K_{ab}^{\phantom{ab}c} \equiv \frac{1}{2} g^{cd} \l( T_{adb} + T_{bda} - T_{abd} \r)  \ ,
\label{contorsion2} 
\end{equation}
and finally, $L^a_{\phantom a b c}$ is defined in terms of the non-metricity tensor, $Q_{abc} \equiv \nabla_a g_{bc}$, as:
\begin{equation}
L^a_{\phantom a b c} \equiv \frac{1}{2} \l( Q^a_{\phantom a cb} - Q_{bc}^{\phantom{bc}a} - Q_{cb}^{\phantom{cb}a} \r) \ .
\label{non-metricity-connection} 
\end{equation}

The GR spacetime is normally taken to be a four dimensional torsionless manifold that satisfies the \emph{metric postulate}:
\begin{equation}
Q_{abc} = 0 \ .
\label{metric_postulate}
\end{equation}
That is to say that both $K^a_{\phantom a b c}$ and $L^a_{\phantom a b c}$ vanish and the connection, now symmetric on the lower indices, reduces to the Levi-Civita connection which is completely defined by the metric tensor.

The curvature of a region of a manifold is described by the change of the order of the double covariant differentiation of a vector field:
\begin{equation}
v_{a;bc} - v_{a;cb} = R^d_{\phantom dabc} v_d \ ,
\label{vector_curv}
\end{equation}
where
\begin{equation}
R^d_{\phantom dabc} \equiv \Gamma^d_{\phantom dac,b} - \Gamma^d_{\phantom dab,c} + \Gamma^e_{\phantom eac} \Gamma^d_{\phantom deb} - \Gamma^e_{\phantom eab} \Gamma^d_{\phantom dec} \ ,
\end{equation}
is called the \emph{Riemann tensor} or the \emph{curvature tensor}. It obeys the following symmetries:
\begin{align}
& R_{abcd} = R_{cdab} \ , &&\mathrm{symmetry\ over\ the\ first\ and\ last\ pair\ of\ indices}, \label{Riemann_sym}\\
& R_{abcd} = R_{[ab][cd]} \ , &&\mathrm{antisymmetry\ in\ the\ first\ and\ last\ pair\ of\ indices}, \label{Riemann_anti_sym}\\
& R_{a[bcd]} = 0 \ , &&\mathrm{the\ cyclic\ identities}, \label{cyclic_id} \\
& R_{[ab|cd|;e]} = 0 \ , &&\mathrm{the\ Bianchi\ identities}. \label{bianchi_id} 
\end{align}
The last relation follows from the cyclic identities which in turn have been derived from the symmetry relations \eqref{Riemann_sym} and \eqref{Riemann_anti_sym}.

The Ricci identities (Eq. \ref{vector_curv}) and the Bianchi identities (Eq. \ref{bianchi_id}) are geometrical results of extreme importance in cosmology. They constitute the starting point in deriving the propagation equations used in the next chapter \cinc \footnote[\countval]{see Appendix \ref{Propagation_eqs_deriv}.}.

Contraction over the first and third indices of the Riemann tensor yields the \emph{Ricci tensor}:
\begin{equation}
R_{ab} \equiv g^{cd} R_{cadb} = R^c_{\phantom cacb} \ .
\end{equation}
Another contraction over the two indices of the Ricci tensor gives the \emph{Ricci scalar}:
\begin{equation}
R \equiv g^{ab} R_{ab} = R^a{}_a \ ,
\end{equation}
defined at every point of the manifold.

It is worth ending this section by stating a relation of great importance in relativity. From the Bianchi identities (Eq. \ref{bianchi_id}), and using the antisymmetry relation (Eq. \ref{Riemann_anti_sym}) it follows that:
\begin{equation}
\l( R_{ab} - \frac{1}{2} g_{ab} R \r)^{;b} = 0 \ .
\end{equation}
The Einstein tensor, $G_{ab} \equiv R_{ab} - \frac{1}{2} g_{ab} R$, is therefore divergence free \cite{Hobson}.

\section{Einstein's Theory of General Relativity}

To model the universe, one needs theories that describe the fundamental interactions. On the large scale, the universe is governed by gravity, the most familiar interaction of them all; yet the least understood. The most successful gravitation theory is Einstein's theory of general relativity, which survived many tests at the level of the solar system \cite{Uzan:2010ri}. 

There are many formalisms of GR; reviewed in \cite{Peldan:1993hi}. The most common ones are the metric, the Palatini, and the metric affine formalisms. 

\subsection{The Metric Formalism of GR}

The simplest choice for the gravitational Lagrangian is the Einstein-Hilbert in-vacuo Lagrangian \cite{Hobson}:
\begin{equation}
\mathcal L_{EH} = R \ .
\end{equation}
The Ricci scalar depends only on the metric tensor and its derivatives of second order at most. Including the cosmological constant $\Lambda$, the gravitational Lagrangian has the form:
\begin{equation}
\mathcal L_{GR} = R(g_{ab}) - 2 \Lambda \ .
\label{GLDE} 
\end{equation}
The GR action is therefore:  
\begin{equation}
S_{GR} = \int_\mathcal{V} \mathcal{L}_{GR} \sqrt{-g} \ \text d^4x \ ,
\label{actionEHDE} 
\end{equation}
where $g$ is the determinant of the metric tensor $g^{ab}$. 

We consider the total action being the sum of GR and the matter field actions:
\begin{equation}
S_{tot} = S_{GR} + S_m \ ,
\end{equation}
where $S_m \equiv \int_\mathcal{V} \mathcal L_m(g_{ab},\Psi_m) \sqrt{-g} \text d^4x$ with $\mathcal L_m(g_{ab},\Psi_m)$ is the matter Lagrangian depending on the metric and the matter field $\Psi_m$. 

Varying the action over some volume $\mathcal{V}$ with respect to the metric leads to the Einstein field equations (EFE) with a cosmological constant:
\begin{equation}
R_{ab} - \frac{1}{2} \ g_{ab} \ R + \ g_{ab} \ \Lambda = T_{ab} \ ,
\label{EFE} 
\end{equation}
where $T_{ab}=-\frac{2}{\sqrt{-g}} \frac{\delta \mathcal{L}_m}{\delta g^{ab}}$ \cite{Hobson,Uzan}. In the derivation of equations \eqref{EFE} from equations \eqref{GLDE}, the surface term does not vanish by mere boundary fixing of the metric, 
$$\delta g_{ab}\biggl|_{\partial \mathcal{V}} = 0 \ ,$$
where $\partial \mathcal{V}$ denotes the region boundary. Fortunately, this surface term is a total variation. Thus it can be cancelled out by adding, to the action, a total divergence; the Gibbons-Hawking-York (GHY) surface term \cite{York:1972sj,Gibbons:1976ue}. 

\subsection{The Palatini Formalism of GR}
An insightful approach to GR is the Palatini formalism which presumes no dependence between the metric and the connection fields \cite{Hobson}. The Einstein-Hilbert action depends only on the dynamical fields i.e. the connection:
\begin{equation}
S_{Palatini} = \int_\mathcal{V} \l[ \mathcal{R} \l( \Gamma \r) - 2 \Lambda \r] \sqrt{-g} \text d^4x + S_m(g_{ab},\Psi_m)  \ ,
\label{actionPalatiniGR} 
\end{equation}
where $\mathcal{R} \l( \Gamma \r)$ denotes the Ricci scalar derived from the connection. Variation of the action \eqref{actionPalatiniGR} with respect to the connection yields the metric postulate:
\begin{equation}
g_{ab;c}=0 \Longleftrightarrow g_{ab,c} = g_{bd} \Gamma^d_{\phantom dac} + g_{ad}\Gamma^d_{\phantom dbc} \ ,
\label{metricpostulate} 
\end{equation}
which in turn leads, via cyclic permutation of the free indices, to the equivalence of the general affine and the Levi-Civita connections in GR. Remarkably, the metric compatibility of the connection derives naturally from the action variation with respect to the connection without a priori assumption. Variation with respect to the metric thus gives the EFE (Eq. \ref{EFE}) \cite{Hobson,Clifton:2011jh}. 

\subsection{The Metric-Affine Formalism of GR}
The metric-affine gravity is another interesting approach which is in fact a generalization of both the metric and the Palatini formalisms above. Here, the matter Lagrangian couples not only to the matter field and the metric but also to the connection \cite{Clifton:2011jh}. The metric compatibility of the latter has not yet been established. The metric-affine action is:
\begin{equation}
S_{m.affine} = \int_\mathcal{V} \l[ \mathcal{R} \l( \Gamma \r) - 2 \Lambda + \mathcal L_m(g_{ab},\Gamma,\Psi_m) \r] \sqrt{-g} \text d^4x \ ,
\label{actionMAGR} 
\end{equation}
which, when varied with respect to the connection gives:
\begin{equation}
S^c_{\phantom cab} + 2 \delta^c_{\phantom c[a} S^d_{\phantom db]d} + \delta^c_{\phantom c[a} Q_{b]} - \delta^c_{\phantom c[a} \bar Q_{d]b}^{\phantom{d]b}d} = \frac{g_{bd}}{\sqrt{-g}} \frac{\delta \mathcal{L}_m}{\delta \Gamma^c_{\phantom cad}} \ ,
\label{MAFE} 
\end{equation}
with $S^c_{\phantom cab} \equiv \Gamma^c_{\phantom c[ab]}$, $Q_a \equiv \frac{1}{4} Q_{ab}^{\phantom{ab}b}$, and $\bar Q_{abc} \equiv Q_{abc} - Q_a g_{bc}$. The left hand side of the field equations \eqref{MAFE} is invariant under projective transformations:
\begin{equation}
\Gamma^a_{\phantom abc} \rightarrow \Gamma^a_{\phantom abc} + \delta^a_{\phantom a b} \xi_c \ ,
\label{projective_transform}
\end{equation}
for an arbitrary vector field $\xi_c$. The matter term on the right hand side does not have to be invariant under these transformations. The non-metricity and the torsion are therefore required to vanish for equations \eqref{MAFE} to be consistent. As well, the metric affine action naturally imposes vanishing torsion and non-metricity and thus the metric compatibility of the connection. Variation with respect to the metric then leads to the EFE \cite{Vitagliano:2010sr,Sotiriou:2006qn}.

\section{The Cosmological Principle}
\label{cosmo_princip}
\emph{The cosmological principle} states that there is no favoured location in the Universe. It rises from \emph{isotropy} and the assumption of \emph{homogeneity}. The Universe appears to be isotropic since, on large scales, one starts to see smooth structure in every direction. In addition, the CMB has the same temperature over the whole sky to high accuracy; one part in a hundred thousand. Homogeneity, on the other hand, cannot be directly observed. Verifying isotropy in another point in space (2 other points in spherical geometries) would prove homogeneity. Assuming \emph{The Copernican principle} implies homogeneity \cite{Clarkson:2010uz,Peebles:1998yv,Hobson,Maroto:2004pd}.

\section{The Friedmann-Lema\^itre-Robertson-Walker Metric}
The EFE (Eq. \ref{EFE}) are non-linear second order differential equations. A few exact solutions have been produced so far \cite{stephani2003exact}. One of them is the work of Friedmann and Lema\^itre based on the EFE \cite{Friedmann:1922,Lemaitre:1933gd}. Later, Robertson and Walker \cite{Robertson:1933zz,Walker:1933MNRAS..94..159W} developed a cosmological model by considering the Copernican principle and making use of symmetries. Their result is a geometrical one and is not founded on a specific field equation. The derivation is covered exhaustively in most standard text books. 
The Friedmann-Lema\^itre-Robertson-Walker (FLRW) line element in 4-dimensional spacetime is:
\begin{equation}
\text ds^2 = - \text dt^2 + a^2(t) \text d\Sigma^2 \ ,
\label{FLRWmetric}
\end{equation}
where $\text dt^2$ and $d\Sigma^2$ are the cosmic time and the spatial intervals \cite{Hobson}. As its name suggests, the scale factor $a(t)$ is a measure of the length scale of the universe. It is a function of time and is normalised to a today value of one; i.e. $a(t=\mathrm{now})=1$. In polar coordinates: 
$$\text d\Sigma^2 = \frac{\text dr^2}{1+Kr^2} + r^2 \l( \text d\theta^2 + sin^2\theta \  \text d\phi^2 \r) \ ,$$
where $K\in \{ -1,0,1 \}$ represents open, flat, and closed geometries respectively. It is common to write the FLRW metric in terms of the conformal time interval, $\text d\eta \equiv \frac{\text dt}{a}$, giving:
\begin{equation}
\text ds^2 = a^2(\eta) \l[ - \text d\eta^2 + \frac{\text dr^2}{1+Kr^2} + r^2 \l( \text d\theta^2 + sin^2\theta \ \text d\phi^2 \r) \r].
\end{equation}

Using the FLRW metric above, an alternate form of the EFE:
\begin{equation}
R_{ab} = T_{ab} + \frac{1}{2} \ g_{ab} \ R + \ g_{ab} \ \Lambda \ ,
\label{EFEalternate} 
\end{equation}
and the perfect fluid model for the EMT:
\begin{equation}
T_{ab} = (\rho -p) u_a u_b + p g_{ab} \ ,
\label{EMTperfect} 
\end{equation}
one gets the cosmological field equations; also known as the Friedmann--Lema\^itre equations \cite{Hobson}:
\begin{subequations}
\label{friedmann_eqs}
\begin{eqnarray} 
&& H^2 = \frac{1}{3} \rho + \frac{1}{3} \Lambda - K \label{friedmann_dot} \ ,\\
&& \frac{\ddot a}{a} = - \frac{1}{6} \l(\rho + 3 p \r) + \frac{1}{3} \Lambda  \ , \label{friedmann_ddot}
\end{eqnarray}
\end{subequations}
where the dot denotes differentiation with respect to the cosmic time $t$ and $H \equiv \frac{\dot a}{a}$ is the local Hubble parameter which determines the expansion rate of the distance between neighbouring points in space. 

On the other hand, the energy conservation condition, $\nabla^a T_{ab} = 0$, for equation \eqref{EMTperfect} leads to the equation of motion (EoM):
\begin{equation}
\dot \rho + 3 H \l( \rho + p \r) = 0 \ .
\label{Econserv}
\end{equation}
Only two of equations \eqref{friedmann_dot}, \eqref{friedmann_ddot}, and \eqref{Econserv} are independent. The third is inferred easily.

To close the system, a relation between the pressure and the density of the fluid is needed. For barotropic fluids with a linear relation between $p$ and $\rho$, the equation of state (EoS) of the fluid plays just this role: 
\begin{equation}
p = \omega \rho \ ,
\label{EoS}
\end{equation}
where $\omega$ is called the EoS parameter and is usually assumed to be constant in time for standard fluids. However, some cosmological models may involve exotic fluids with $\omega=\omega(t)$ \cite{Hobson}. 

Direct solution of equation \eqref{Econserv} leads to the density evolution equation of every conserved matter species $i$:
\begin{equation}
\rho_i(t) = \rho_{i,0} \l[ a(t) \r]^{-3(1+\omega_i)} \ ,
\label{density_evolution}
\end{equation}
where $\rho_{i,0}$ is the present day density of species $i$ and the present value of the scale factor has been normalised to unity. 

The following table states the EoS parameter and the density evolution of some common fluids:

\begin{center}
\begin{tabular}[!htpd]{c|c|c}
Fluid & $\omega$ & $\rho(a)$\\ \hline
Dust & $0$ & $a^{-3}$ \\
Radiation & $\frac{1}{3}$ & $a^{-4}$ \\
The cosmological constant & $-1$ & constant \\
\end{tabular}
\end{center}

Finally, we finish this section by introducing a useful quantity. The density parameter is defined as:
\begin{equation}
\Omega_i(t) \equiv \frac{\rho_i(t)}{3 H^2(t)}  \ .
\label{density_params}
\end{equation}
Rewriting the Friedmann equation above in terms of the density parameters yields an important relation in cosmology:
\begin{equation}
\sum_i \Omega_i = 1 \ .
\end{equation}

\section{The Concordance Model}
The universe we live in is not as smooth as an FLRW universe would be; but rather \emph{perturbed}. $\Lambda$CDM, is the best fit model to observations cosmologists have. To achieve this status, 95\% of the content of the universe needs to be \emph{dark} \cite{Clifton:2011jh}. Indeed, $\Lambda$CDM claims that observable baryonic matter constitutes only 4 to 5\% of the total energy density of the universe. Another 25\% is in the form of non-relativistic \emph{dark matter} interacting only gravitationally. Dark matter was, for instance, strongly suggested by rotational curves of disk galaxies \cite{Rubin:1985ze}. The surprising discovery of the acceleration of the expansion of the universe through distant type Ia supernovae measurements \cite{Perlmutter:1997zf,Riess:1998cb,Perlmutter:1998np} imposed that the remaining 70\% of the energy density in the $\Lambda$CDM picture has to be some unknown form of energy with anti-gravitational properties. This became known as \emph{dark energy} \cinc \footnote[\countval]{The $\Lambda$CDM model is based on a number of assumptions such as CDM and inflation. Dark energy is inferred by fitting the model to the available data.}.
\section{Beyond the Concordance Model}
The concordance model suffers from a few problems (the cusp-core problem \cite{deBlok:2009sp,Li:2009mp}, the missing satellite problem \cite{Bullock:2010uy}, among others) out of which that of the cosmological constant is the most severe \cite{Clifton:2011jh}. The value of the vacuum energy density, $\rho_\Lambda$, calculated at near the Planck scales is 120 orders of magnitude larger than the upper bound set by cosmological observations. Proposed mechanisms (such as super-symmetry theories) to reconcile these results have to be accurate to within 120 decimal places causing a \emph{fine tuning problem}. In addition, the current densities of matter and dark energy are of of the same order of magnitude, $\rho_{\Lambda,0} \approx \rho_{m,0}$. This is the so called \emph{coincidence problem} \cite{Weinberg:2000yb,Clifton:2011jh}.

The problems with the $\Lambda$CDM model mainly reflect the shortcomings of the classical theory of general relativity. This created both the need and the opportunity to investigate more general theories of gravity. The idea is almost as old as GR itself. Barely a few years after Einstein published his paper on GR \cite{Einstein:1915by}, there were already suggestions of modifying the Einstein-Hilbert action by introducing higher order terms \cite{Weyl:1919fi,Eddington1923}. The first requirement of any gravitation theory is satisfying the uncompromising condition of fitting astrophysical and cosmological observations. \emph{Extended theories of gravity} strive to build up on the success of GR by introducing corrections either through coupling geometry to a scalar field in which case the result is a scalar-tensor theory of gravity, or by adding higher order curvature invariants to the gravitational Lagrangian; this accordingly yields, in general, to higher order field equations \cite{Capozziello:2007ec}. So while attempting to explain the universe at high energy regimes, ETGs should absolutely reduce to GR in the weak field limit.



\chapter{The Covariant Approach to Perturbations in the CMB} 
\label{cmb}
%
%
\ifpdf
    \graphicspath{{3/figures/PNG/}{3/figures/PDF/}{3/figures/}}
\else
    \graphicspath{{3/figures/EPS/}{3/figures/}}
\fi
%
%
%
\section{The Observables}
For every point in space $\vec x$, time $\eta$, and photon incidence direction $\vec e$, the perturbations in the temperature field can be written as:
\begin{equation}
T \l( \vec x,\eta,\vec e \r) = T(\eta) \l[ 1 + \tilde \Theta \l( \vec x,\eta,\vec e \r) \r] \ ,
\label{tempfield}
\end{equation}
where $\tilde \Theta \equiv \frac{\delta T}{T}$. It directly follows that $T \l( \vec x,\eta,\vec e \r) = T(\eta)$ if the Copernican principle is assumed \cite{Uzan}. Although $\tilde \Theta \l( \vec x,\eta,\vec e \r)$ characterises $T \l( \vec x,\eta,\vec e \r)$ at every point in spacetime, it can be observed only locally, i.e. at $(\vec x_0, \eta_0)$. 

$\tilde \Theta \l( \vec x_0,\eta_0,\vec e \r)$ is stochastic and thus not useful per se, since the mean is zero. A good statistical tool is the \emph{correlation function}:
\begin{equation}
C(\vartheta) = \la \tilde \Theta \l( \vec x_0,\eta_0,\vec e_1 \r) \tilde \Theta \l( \vec x_0,\eta_0,\vec e_2 \r) \ra  \ ,
\label{CF}
\end{equation}
where $\vartheta = \arccos(\vec e_1 \cdot \vec e_2)$ is the relative angle between the photon incidence directions and the angle brackets denote the average over the whole distribution. 

The perturbation in the temperature field can be expanded in terms of spherical harmonics as:
\begin{equation}
\tilde \Theta \l( \vec x_0,\eta_0,\vec e \r) = \sum_{l=1}^\infty \sum_{m=-l}^l a_{lm}(\vec x_0,\eta_0) Y_{lm}(\vec e) \ .
\label{tempsh}
\end{equation}
Again, the coefficients $a_{lm}$ do not provide any predictions due to the randomness of the temperature fluctuations \cite{Dodelson}. The mean of the $a_{lm}$ coefficients is zero. On the other hand, the quantity:
\begin{equation}
C_l^{TT} = \la a^T_{lm}(\vec x_0,\eta_0) \, a^{* T}_{lm}(\vec x_0,\eta_0) \ra
\end{equation}
is the angular power spectrum and measures the variance of the temperature fluctuations at the angular scales corresponding to multipole $l$, approximately $\pi/\vartheta$ \cite{Uzan,Dodelson}.

Using the normalisation of the spherical harmonics \cite{riley2006mathematical}:
\begin{equation}
\int Y_{lm}(\vec e) Y^*_{lm}(\vec e) \text d\Omega = 1
\end{equation}
and equation \eqref{tempsh}, it can be shown that the correlation function in equation \eqref{CF} can also be expressed in terms of Legendre polynomials as:
\begin{equation}
C(\vartheta) = \sum_l \frac{2 l + 1}{4 \pi} \Cl \Pl \l( \vec e_1 \cdot \vec e_2 \r) \ .
\end{equation}
The $C_l$ coefficients then provide a good characterization of the fluctuations in the CMB on different scales. 

So far, we have only considered the correlation function in temperature. Similarly, other observables may be obtained by:
\begin{equation}
C_l^{X_1 X_2} = \la a^{X_1}_{lm}(\vec x_0,\eta_0) \, a^{* X_2}_{lm}(\vec x_0,\eta_0) \ra \ ,
\label{Power_Spectrum_Cl}
\end{equation}
where $X_1,X_2 \in \{ T, E, B \}$ \cite{Hu:2002aa,Baskaran:2006qs}. The scalar $E$ and pseudo-scalar $B$ represent the $E$-mode and $B$-mode polarisations of the CMB that we briefly describe below.

The polarisation of an electromagnetic wave can be described by the Stokes parameters $I, \ Q, \ U, \ \text{and} \ V$ encoded in the coherence matrix:
\begin{equation}
C = \frac{1}{2}\left( \begin{array}{cc}
\langle{}I+Q\rangle & \langle{}U-iV\rangle \\
\langle{}U+iV\rangle & \langle{}I-Q\rangle \end{array} \right) \ .
\end{equation}
$I$ is the total intensity of the electric component of the electromagnetic wave. $Q$ and $U$ respectively represent the horizontal/vertical and the $\pm 45^\circ$ linear polarisations of the electric vector $\vec E$. The last parameter, $V$, is the left and right hand circular polarisations of $\vec E$. The Stokes parameters satisfy the inequality:
\begin{equation}
I^2 \geq Q^2 + U^2 + V^2 \ ,
\end{equation}
where equality happens if an only if the wave is fully polarised. For the CMB radiation, the parameter $V$ is expected to vanish as it cannot be generated via Thompson scattering \cite{Zaldarriaga:1996xe}.

Since $Q\pm iU$ is a spin 2 object, it can be expressed in terms of spin 2 spherical harmonics:
\begin{equation}
(Q\pm iU)(\hat{n})=\sum_{l\geq2,|m|\leq l}a_{\pm2lm}\ {}_{\pm2}Y^m_l(\hat{\vec n}) \ .
\label{spin2spherical_harmonics}
\end{equation}
The Stokes parameters, unlike the temperature, are not invariant under rotation transformations; which is inconvenient for computing the CMB power spectra. This can be overcome by using the spin raising and lowering operators, $\edth$ and $\baredth$, to obtain spin zero quantities that are rotationally invariant (See \cite{Zaldarriaga:1996xe} for more details). Applying $\edth^2$ and $\baredth^2$ on  equation \eqref{spin2spherical_harmonics} gives:
\begin{subequations}
\begin{eqnarray}
\baredth^2(Q+iU)(\hat{\bi{n}})&=& \sum_{lm} \left[{(l+2)! \over (l-2)!}\right]^{1/2} a_{2,lm}Y_{lm}(\hat{\bi{n}}) \ , \\
\edth^2(Q-iU)(\hat{\bi{n}})&=&\sum_{lm} \left[{(l+2)! \over (l-2)!}\right]^{1/2} a_{-2,lm}Y_{lm}(\hat{\bi{n}}) \ ,
\end{eqnarray}
\end{subequations}
allowing one to define two scalar (invariants) quantities:
\begin{subequations}
\begin{eqnarray}
E(\hat{{\bi n}}) &\equiv& -{1\over 2}\left[\baredth^2(Q+iU)+\edth^2(Q-iU)\right] \nonumber \\
	&=&\sum_{lm}\left[{(l+2)! \over (l-2)!}\right]^{1/2} a_{E,lm}Y_{lm}(\hat{{\bi n}}) \ ,  \label{Eexpansions}\\  
B(\hat{\bi n})&\equiv&{i\over 2} \left[\baredth^2(Q+iU)-\edth^2(Q-iU)\right] \nonumber \\
	&=&\sum_{lm}\left[{(l+2)! \over (l-2)!}\right]^{1/2} a_{B,lm}Y_{lm}(\hat{\bi n}) \ .  \label{Bexpansions}
\end{eqnarray}
\label{EBexpansions}
\end{subequations}
where $a_{E,lm}$ and $a_{B,lm}$ are linear combinations of $a_{2,lm}$ and $a_{-2,lm}$:
\begin{eqnarray}
a_{E,lm} = - \ \frac{a_{2,lm}+a_{-2,lm}}{2} \ , \\ 
a_{B,lm} = i \ \frac{a_{2,lm}-a_{-2,lm}}{2} \ .
\label{aeb}
\end{eqnarray}
Parity transformations keep $a_{E,lm}$ unchanged while it changes the sign of $a_{B,lm}$ \cite{Newman:1966ub}. The power spectra involving $E$ and $B$ can be obtained using equation \eqref{Power_Spectrum_Cl}. 

In cosmology, the cross correlations $C_l^{TT}, \ C_l^{EE}, \ C_l^{TE}, \ \text{and} \ C_l^{BB}$ fully characterise the statistics of the CMB perturbations. The cross correlation $C_l^{TB}$ and $C_l^{EB}$ are zero due to the negative parity of $B$ \cite{Zaldarriaga:2003bb}.

%


\section{Scalar-Vector-Tensor Decomposition of the Perturbation Metric}
\label{svt_decomp}
The scalar-vector-tensor decomposition is based on the fact that any 3-vector can be expressed as the sum of its rotational and irrotational parts \cite{Bertschinger:2001is,Clarkson:2011td}:
\begin{equation}
v_i = v_i^\parallel + v_i^\perp \ ,
\label{vec_decomp}
\end{equation}
such that $\D^i v_i^\perp = \curl v_i^\parallel = 0$. The irrotational, or longitudinal, vector can be written as the gradient of a scalar:
\begin{equation}
v_i^\parallel = \D_i\phi_v \ ,
\end{equation}
while the rotational, or transverse, part cannot be obtained from a scalar.


Similarly, a rank two tensor can be decomposed into a doubly longitudinal, a singly longitudinal, and a doubly transverse tensor \cite{Bertschinger:2001is,Clarkson:2011td}:
\begin{equation}
S_{ij}=S_{ij}^\parallel+S_{ij}^\perp+S_{ij}^{\rm T} \ ,
\label{tensor_decomp}
\end{equation}
such that 
\begin{equation}
\D^j S_{ij} = \D^j S_{ij}^\parallel + \D^j S_{ij}^\perp \ .
\label{tensor_div}
\end{equation}
The divergence of the doubly transverse component is zero. For a symmetric and trace free tensor $S_{ij}$, the doubly longitudinal part can be
obtained from the double gradient of a scalar $\phi_S$:
\begin{equation}
  S_{ij}^\parallel = \l( \D_i \D_j - \frac{1}{3} h_{ij} \D^2 \r) \phi_S \ ,
  \label{longtens}
\end{equation}
while the singly longitudinal part can be obtained from the gradient of transverse vector $S_j^\perp$:
\begin{equation}
S_{ij}^\perp = 2 \D_{(i} S_{j)}^\perp \ .
\end{equation}

The perturbed metric, in linear perturbation theory, can be regarded as a (small) symmetric tensor $\delta g_{ab}$ residing on the FLRW metric \cite{Uzan}:
\begin{equation}
g_{ab} = g_{ab}^{\mathrm{FLRW}} + \delta g_{ab} \ ,
\end{equation}
so the the perturbed FLRW line element is \cite{Bertschinger:2001is}:
\begin{equation}
\text ds^2 = a^2(\eta) \l[ - \text d\eta^2 + \gamma_{ij} \ \text dx^i \text dx^j + \xi_{ab} \ \text dx^a \text dx^b \r] \ ,
\end{equation}
where $\gamma_{ij}$ is the 3-metric and $\xi_{ab} \equiv \frac{\delta g_{ab}}{a^2(\eta)}$ has the components:
\begin{equation}
  \xi_{00} \equiv - 2 \psi\ , \quad
  \xi_{0i}\equiv v_i\ , \quad
  \xi_{ij} \equiv 2 \l( \phi \gamma_{ij} + S_{ij} \r) \ ,
  \label{pertur_tensor}
\end{equation}
with $S^i_{\phantom ii}=0$. The trace of $\xi_{ij}$ has been absorbed into the scalar $\phi$.

The vector part of $\xi_{0i}$ cannot be obtained from the derivatives of a scalar as well as the tensor part of $\xi_{ij}$ cannot be obtained from the derivatives of scalars and vectors. The scalar modes behave like spin $0$ fields under spatial rotations, the vector modes like spin  $1$, and the tensor modes like spin $2$.
In cosmology, the scalar modes correspond to density perturbations. The vector and tensor modes do not affect the density and thus are not important to the structure formation although they do distort the CMB. The vector modes, corresponding to gravitomagnetism, decay with the expansion of the Universe and, therefore, are not important to the evolution of perturbations \cite{Mukhanov:1990me}. Finally, the tensor modes are the source of gravitational radiation. $\xi_{ij}$ possesses two degrees of freedom representing the two polarisations $\xi_{xx}$ and $\xi_{yy}$, with $\xi_{xy}=\xi_{yx}$ \cite{Bertschinger:2001is}.

\section{The Covariant and Gauge Invariant Approach to Perturbations}
\subsection{Gauge Invariance}
If a quantity $Q$ is expressed as the corresponding background quantity plus a small perturbation, i.e.:
\begin{equation}
Q = Q_0 + \delta Q \ ,
\label{quantity+pertubations}
\end{equation}
the \emph{gauge} transformation $Q'$ of $Q$ along an infinitesimal vector field $\boldsymbol \xi$  is written as:
\begin{equation}
Q \to Q' = Q + \pounds_{\boldsymbol \xi} Q_0 \ ,
\label{gauge_transform}
\end{equation}
where the operator $\pounds_{\boldsymbol \xi}$ is the Lie derivative defined as:
\begin{equation}
\pounds_{\boldsymbol \xi} T_{a_1 \ldots a_p}^{b_1 \ldots b_q} \equiv \xi^c \p_c T_{a_1 \ldots a_p}^{b_1 \ldots b_q} - \sum_{i=1}^p T_{a_1 \ldots a_p}^{b_1 \ldots c \ldots b_q} \p_c \xi^{b_i} + \sum_{j=1}^q T_{a_1 \ldots c \ldots a_p}^{b_1 \ldots b_q} \p_{a_i} \xi^{c} \ ,
\label{Lie_deriv}
\end{equation}
%
with $T_{a_1 \ldots a_p}^{b_1 \ldots b_q}$ being general a rank $p+q$ tensor. Equations \eqref{quantity+pertubations} and  \eqref{gauge_transform} imply:
\begin{equation}
\delta Q ' = \delta Q + \pounds_{\boldsymbol \xi} Q_0 \ . 
\end{equation}
to first order in perturbations. 

It is deduced that only quantities for which:
\begin{equation}
\pounds_{\boldsymbol \xi} Q_0 = 0 \ , \qquad \forall \boldsymbol \xi \ ,
\end{equation}
are gauge invariant (GI) \cite{Uzan,Bruni:1992dg}. This is the Stewart--Walker Lemma \cite{Stewart:1974uz}. It means that perturbations to a background quantity are GI if and only if one of the following conditions holds:
\begin{enumerate}
\item $Q_0 = 0$,
\item $Q_0$ is a constant scalar,
\item $Q_0$ is a linear combination of products of Kronecker deltas.
\end{enumerate}

The metric perturbations $\xi_{ab}$ presented in the previous section is a symmetric tensor with ten degrees of freedom of which six only are physical \cite{Bertschinger:2001is}. The other four are gauge dependent. As a matter of fact, the metric approach has the disadvantage that the metric tensor, or the perturbations thereof, are not physically meaningful \cite{Hawking:1966qi}. A perturbed quantity can be physically interpreted only after a map between the perturbed universe and the unperturbed background (usually FLRW) has been specified; i.e. a fixed gauge \cite{Bruni:1992dg}. Otherwise, the quantity is arbitrary.

The importance of gauge invariance lies in the fact that observables correspond to GI quantities at first order in perturbations, regardless of whether they are GI at higher orders or not \cite{Bruni:1999et}. A fully GI theory of linear perturbations was presented in \cite{Bardeen:1980kt}. However, the GI variables therein do not have straight forward geometrical interpretation as they are constructed in terms of gauge dependent variables. This is mainly because the term $\delta Q$ in equation \eqref{quantity+pertubations} is not a tensor and its meaning, therefore, depends on the chosen coordinate system \cite{Bruni:1992dg}. 

Another approach that is both covariant and GI was suggested by \cite{Hawking:1966qi} and developed further in \cite{Ellis:1989jt,Bruni:1992dg}. It is basically based on curvature variables as it is the second derivatives of the metric tensor that can be observed, rather than the metric tensor itself.

\subsection{The $1+3$ Formalism of the Covariant Approach}
The $1+3$ formalism provides an insightful approach for dealing with the dynamics of cosmological models \cite{Challinor:1999xz}. The following is based on \cite{Ellis:1989jt} \cite{Betschart:2005PhD} \cite[pp 34--35]{Uzan}.
\subsubsection{The Time and Space Projection Tensors}
Spacetime is sliced into constant time hyper-surfaces with respect to fundamental observers with 4-velocity:
\begin{equation}
u^a \equiv \frac{dx^a}{d\tau} \ ,
\end{equation}
where $\tau$ is the proper time along the observers' world lines. The definition directly implies that:
\begin{equation}
u^a u_a = -1 \ ,
\end{equation}
which means that the velocity vector is time-like. The 4-metric tensor may then be decomposed into time and rest-space symmetric projection tensors:
\begin{equation}
g_{ab} = - u_a u_b + h_{ab} \ .
\label{metric_decomp}
\end{equation}
The tensor $U_{ab} \equiv - u_a u_b$ projects on the parallel to $u_a$. It can directly be verified that:
\begin{equation}
U_{ab} u^a = u_b, \qquad U^a{}_b U^b{}_c = U^a{}_c, \qquad U^a{}_a = 1 \ .
\label{time_projection_properties}
\end{equation}
$h_{ab} $ is the projection tensor on the constant time hyper-surface. From equations \eqref{metric_decomp} and \eqref{time_projection_properties}, we have:
\begin{equation}
h_{ab} u^a = 0, \qquad h^a{}_b h^b{}_c = h^a{}_c, \qquad h^a{}_a = 3 \ .
\label{space_projection_properties}
\end{equation}

If the velocity vector $u_a$ is orthogonal to the constant time hyper-surface, the space projection tensor $h_{ab}$ would also be the metric tensor on the rest-space.

\subsubsection{Derivatives}
The projected, totally antisymmetric tensor on the rest space of a comoving observer is defined as:
\begin{equation}
\epsilon_{abc} \equiv \eta_{abcd}u^d \ ,
\end{equation}
where $\eta_{abcd}$ is the totally antisymmetric tensor on spacetime given by:
\begin{equation}
\eta^{abcd} =  \eta^{[abcd]}\qquad \text{and} \qquad \eta^{0123} = (-g)^{-\frac{1}{2}} \ .
\end{equation}
In an orthonormal frame, the volume elements $\epsilon_{abc}$ and $\eta_{abcd}$ are alternating quantities \cite{Ellis:2009go}.

Totally antisymmetric tensors of any rank $n$ satisfy the identity \cite{Schutz}:
\begin{equation}
\epsilon^{a_1 a_2 \ldots a_n} \, \epsilon_{b_1 b_2 \ldots b_n} = n! \, \delta^{a_1}_{\ph{a_1} [ b_1} \, \delta^{a_2}_{\ph{a_2} b_2} \ldots \delta^{a_n}_{\ph{a_n} b_n ]} \ ,
\end{equation}
where the square brackets denote anti-symmetry over the enclosed indices. Hence $\epsilon_{abc}$ satisfies:
\begin{subequations}
\begin{eqnarray}
&& \epsilon^{abc} \, \epsilon_{def} = 3! \, h^{[a}{}_d \, h^b{}_e \, \delta^{c]}{}_f \ , \\
&& \epsilon^{abc} \, \epsilon_{cef} = 2! \, h^{[a}{}_e \, h^{b]}{}_f \ , \\
&& \epsilon^{abc} \, \epsilon_{bcf} = 2! \, h^a{}_f \ , \\
&& \epsilon^{abc} \, \epsilon_{abc} = 3! \ .
\end{eqnarray}
\end{subequations}

The projected covariant derivative, the time derivative, and the generalized 3-dimensional curl of tensors are, respectively, defined as:
\begin{align}
\D_a T_{b\dots}{}^{c\dots} & \equiv h^d_a h^e_b \dots h^c_f \dots \nabla_d T_{e\dots}{}^{f\dots} \ , \\
\dot T_{b\dots}{}^{c\dots} & \equiv  u^a \nabla_a T_{b\dots}{}^{c\dots} \ , \\
\curl T_{ab\dots c} & \equiv \epsilon_{d e (a} \D^d {T_{b\dots c)}}^e \ ,
\end{align}
where the parentheses denote symmetry over the enclosed indices. If ${T_{ab\dots c}}$ is a projected symmetric trace free (PSTF) tensor, $\curl T_{ab\dots c}$ is also PSTF \cite{Challinor:1999xz}. 
\subsubsection{Kinematics}
The kinematics of the fluid are obtained from the decomposition of the covariant derivative of the 4-velocity. A first decomposition into spatial and temporal components:
\begin{equation}
u_{a;b} = \D_b u_a - u_b A_a \ ,
\label{velocity_decomp_spacetime}
\end{equation}
introduces
\begin{equation}
A_a \equiv \dot u_a \ ,
\label{acceleration_def}
\end{equation}
the acceleration along the flow lines. The spatial part $\D_b u_a$ is further decomposed into its trace, symmetric trace free, and antisymmetric parts:
\begin{equation}
\D_b u_a =  \frac{1}{3} h_{(ab)} \D^c u_c + \D_{\la b} u_{a \ra} + \D_{[b}u_{a]} \ ,
\label{velocity_decomp_space}
\end{equation}
where the angle brackets denote the orthogonal projection of the symmetric trace free part of a tensor and defined as:
\begin{equation}
X_{\la ab \ra} \equiv \left[ h_{(a}{}^c \, h_{b)}{}^{d} - \frac{1}{3} \, h_{ab} \, h^{cd} \right] X_{cd}  \ .
\label{anglebracket_def}
\end{equation}

The first two terms in the RHS of equation \eqref{velocity_decomp_space} constitute the expansion tensor which describes the change of the distance between neighbouring particles in a fluid. The isotropic expansion is determined by the volume expansion scalar:
\begin{equation}
\Theta \equiv \D^a u_a = 3H \ .
\label{expansion_def}
\end{equation}
while the anisotropic expansion:
\begin{equation}
\sigma_{ab} =  \sigma_{(ab)} \equiv \D_{\la a} u_{b \ra} \  
\label{shear_def}
\end{equation}
is the shear and it describes volume conserving distortions to the fluid. 

Finally, the last term in equation \eqref{velocity_decomp_space}:
\begin{equation}
\omega_{ab} = \omega_{[ab]} \equiv \D_{[a}u_{b]} \ , 
\label{vorticity_def}
\end{equation}
is the vorticity tensor which describes rigid rotation in the fluid with respect to a local inertial frame \cite{Ellis:2009go}.

Rewriting equation \eqref{velocity_decomp_spacetime}:
\begin{equation}
u_{a;b} = \frac{1}{3} \Theta h_{ab} + \sigma_{ab} + \omega_{ab} - A_a u_b \ ,
\label{velocity_decomp}
\end{equation} 
neatly presents the irreducible decomposition of the covariant derivative of the 4-velocity vector in terms of the expansion scalar, the shear, the vorticity, and the acceleration.

Another useful kinematic quantity is the projected vorticity vector, defined from the vorticity tensor by:
\begin{equation}
\omega_a \equiv \frac{1}{2} \epsilon_{abc} \omega^{bc} = \frac{1}{2} \curl u_a \ .
\label{vorticity_vec}
\end{equation}

Deriving directly from equations \eqref{acceleration_def}, \eqref{expansion_def}, \eqref{shear_def}, \eqref{vorticity_def}, and \eqref{vorticity_vec}, we have:
\begin{align}
& u^a A_a = 0  \ , \\
& u^a \sigma_{ab} = 0, \qquad \qquad h^{ab} \sigma_{ab} = \sigma^a{}_a = 0 \ , \\
& u^a \omega_{ab} = 0, \qquad \qquad h^{ab} \omega_{ab} = \omega^a{}_a = 0 \ ,  \qquad \qquad \omega^a \omega_{ab} = 0 \ ,
\end{align}
which means that the acceleration is a rest space vector and that the shear and vorticity are PSTF. 

$A_a$, $\sigma_{ab}$, and $\omega_a$ characterise anisotropy. while the projected gradient of the expansion scalar, $\D_a \Theta$, characterises inhomogeneity in the expansion. Naturally, all these quantities vanish in an exact FLRW universe \cite{Challinor:1999xz}.
%
%

\subsubsection{Curvature Tensors}
The tidal forces felt by a body moving along a geodesic are described by the Riemann curvature tensor defined in \S \ref{manifolds}. $R_{abcd}$ may be decomposed into its trace and trace-free parts \cite{Ellis:2009go}:
\begin{equation}
R_{ab}{}^{cd} = 2 g_{[a}{}^{[c} \, R_{b]}{}^{d]} - \frac{R}{3} \, g_{[a}{}^{[c} \, g_{b]}{}^{d]} + C_{ab}{}^{cd} \ ,
\label{Riemann_decomp}
\end{equation}
where $C_{abcd}$ is the \emph{Weyl tensor} and describes the distortions due to tidal forces but does not carry any information about the change in the volume. In vacuum, it is the only contribution to the Riemann tensor. $C_{abcd}$ can be decomposed further into a curl-free and a divergence-free PSTF tensors:
 \begin{eqnarray}
E_{ab}  &\equiv&  u^c u^d C_{acbd}  \ , \\
H_{ab}  &\equiv&  \frac{1}{2} \, \epsilon_{acd} \, u^e C_{be}{}^{cd} \ ,
\end{eqnarray}
which are called the \emph{electric} and \emph{magnetic} Weyl tensors, in analogy with the curl-free electric field and divergence-free magnetic field in electromagnetism. $E_{ab}$ is analogous to the traceless tidal tensor defined from the gravitational potential $\Phi$ in Newtonian gravity \cite{Ellis:2009go}:
$$E_{ab}^\text{Newtonian} \equiv \p_a \p_b \Phi - \frac{1}{3} h_{ab} \p^c \p_c \Phi \ . $$
$H_{ab}$, on the other hand, has no Newtonian counterpart thus it is essential for a full description of gravitational waves \cite{Dunsby:1998hd}.

The Weyl tensor can be fully reconstructed from the electric and magnetic Weyl tensors via:
\begin{equation}
C_{abcd}=\left( \eta_{abpq} \eta_{cdrs} + g_{abpq} g_{cdrs} \right) u^p u^r E^{qs} - \left( \eta_{abpq} g_{cdrs} + g_{abpq} \eta_{cdrs} \right) u^p u^r H^{qs} \ .
\label{weyl}
\end{equation}
where
$$g_{abcd} \equiv g_{ac}g_{bd} - g_{ad} g_{bc} = g_{[ab][cd]} = g_{cdab} = -\frac{1}{2}\eta_{ab}^{\phantom{ab}ef} \eta_{efcd} \ , $$
with $g_{a[bcd]} = 0$ \cite{Bertschinger:1993xt}.

\subsubsection{The Energy-Momentum Tensor}
Consider the most general construction of the energy momentum tensor (EMT):
\begin{equation}
T_{ab}^{tot} = \rho u_a u_b + p h_{ab} + 2 q_{(a} u_{b)} + \pi_{ab} \ ,
\label{EMT}
\end{equation}
where $q_a$ and $\pi_{ab}$ are the total energy flux and the total anisotropic stress respectively \cite{Ananda:2007xh}. The following constraints apply:
\begin{equation}
q_a u^a = 0, \qquad \pi_{ab} u^b = 0, \qquad \pi^a{}_a = 0, \qquad \pi_{ab} = \pi_{(ab)} \ .
\label{EMT_constraints}
\end{equation}

The individual components of the EMT can be extracted through the operations:
\begin{subequations}
\label{EMT_decomp_ops}
\begin{align}
\rho &= T_{ab}^{tot} u^a u^b \ ,  \\
p &= \frac{1}{3} T_{ab}^{tot} h^{ab} \ , \\
q_a &= -T_{cd}^{tot} u^c h^d{}_a \ , \\
\pi_{ab} &= T_{\la ab \ra}^{tot} \ .
\end{align}    
\end{subequations}
%

\section{The Evolution of Perturbations in the CMB}

\subsection{The Propagation Equations}
The evolution of perturbations along the flow lines are described by the propagation equations stated in \cite{Ananda:2007xh,Maartens:1998xg}. Here, we require only the linearised form obtained by considering only up to first order departures from FLRW \cite{Challinor:1999xz}. This is done by treating the density ($\rho$), the pressure ($p$), and the expansion scalar ($\Theta$) as zeroth order quantities and the acceleration ($A_a$), the shear ($\sigma_{ab}$), the vorticity ($\omega_{ab}$), the anisotropic stress ($\pi_{ab}$), the energy flux ($q_a$) and the electric and magnetic Weyl tensors ($E_{ab}$ and $H_{ab}$), together with their derivatives, as first order quantities. Then, all the relatively higher order terms are neglected \cite{Ellis:1989jt}.

There are seven propagation equations. In the linearised form, they read:
\begin{eqnarray}
\dot \rho & = & - \l( \rho + p \r) \Theta - \D^a q_a \ , \label{peq1} \\
\dot \Theta & = & - \frac{1}{3} \Theta^2 - \frac{1}{2} \l( \rho + 3p \r) + \D^a A_a \ , \label{peq2} \\
\dot q_a & = & - \frac{4}{3} \Theta q_a - \l( \rho + p \r) A_a - \D_a p - \D^b \pi_{ab} \ , \label{peq3} \\
\dot \omega_a & = & - \frac{2}{3} \Theta \omega_a + \frac{1}{2} \curl A_a \ , \label{peq4} \\
\dot \sigma_{ab} & = & - \frac{2}{3} \Theta \sigma_{ab} - E_{ab} - \frac{1}{2} \pi_{ab} + \D_{\la a} A_{b\ra} \ , \label{peq5}\\
\dot E_{ab} & = & -\Theta E_{ab} + \curl H_{ab} - \frac{1}{2} \l[ (\rho + p)\sigma_{ab} + \D_{\la a} q_{b\ra} + \dot \pi_{ab} + \frac{1}{3} \Theta \pi_{ab} \r] , \label{peq6} \\
\dot H_{ab} & = & - \Theta H_{ab} - \curl E_{ab} - {\frac{1}{2}} \curl \pi_{ab} \ , \label{peq7}
\end{eqnarray}
constrained by the following equations:
\begin{eqnarray}
\D^a \omega_a  =  0 \ , \label{ceq1}\\
\D^a \sigma_{ab} + \curl \omega_b - {\frac{2}{3}}\D_b \Theta + q_b = 0 \ , \label{ceq2}\\
\D^a E_{ab} +  \frac{1}{3} \Theta q_b - \frac{1}{3} \D_b \rho + \frac{1}{2} \D^a \pi_{ab} = 0 \ , \label{ceq3}\\
\D^a H_{ab} + (\rho + p) \omega_b + \frac{1}{2} \curl q_b = 0 \ , \label{ceq4} \\
H_{ab} - \curl \sigma_{ab} + \D_{\la a} \omega_{b \ra} = 0 \ . \label{ceq5}
\end{eqnarray}

The constraints in equations~(\ref{ceq1})--(\ref{ceq5}) are consistent with the linearised propagation equations (\ref{peq1}--\ref{peq7}) \cite{Challinor:1999xz,Bruni:1992dg}. This can be verified by taking the time derivative of a constraint equation. Using the necessary commutation relations and the propagation equations, one should get the same constraint equation making it valid on all constant time hyper-surfaces.

Perturbations in the $1+3$ split can be decomposed into scalar, vector, and tensor modes which evolve independently of each other \cite{Bertschinger:2001is}. Pure tensor perturbations are obtained by ``turning off'' the scalar and vector modes. This implies that the divergence of the electric and magnetic Weyl tensors together with the vorticity and other projected vectors vanish:
\begin{align}
\D^b E_{ab}&=0 \label{tdivE} \ ,\\
\D^b H_{ab}&=0 \label{tdivH} \ ,
\end{align}
\begin{equation}
A_a = 0 \ , \qquad \omega_a = 0 \ , \qquad q_a = 0 \ , \qquad \D_b \rho = 0 \ , \qquad \D_b \Theta = 0 \ .
\label{tcond}
\end{equation}
up to first order \cite{Dunsby:1998hd}

Equations \eqref{tdivE} and \eqref{tdivH} state that the electric and magnetic Weyl tensors are transverse. Indeed, the linearised gravitational waves are described by degrees of freedom of $E_{ab}$ and $H_{ab}$ \cite{Challinor:1999xz}. It is straightforward to verify from equations \eqref{ceq2} and \eqref{ceq3} that the shear and the anisotropic stress are also divergence free:
\begin{align}
\D^b \sigma_{ab}&=0 \ ,\\
\D^b \pi_{ab}&=0 \ ,
\end{align}
and that the only constraint that survives is equation \eqref{ceq5}. It becomes:
\begin{equation}
H_{ab} = \curl \sigma_{ab} \ .
\label{tceq5}
\end{equation}
The propagation equations~(\ref{peq1})--(\ref{peq7}) reduce to:
\begin{eqnarray}
\dot \rho &=& -\Theta \l( \rho + p \r) \ , \label{tpeq1}\\
\dot \Theta &=& - \frac{1}{3} \Theta^2 - \frac{1}{2} \l( \rho + 3 p \r) \ , \label{tpeq2}\\
\dot \sigma_{ab} &=& - \frac{2}{3} \, \Theta \, \sigma_{ab} - E_{ab} + \frac{1}{2} \pi_{ab} \ , \label{tpeq3}\\
\dot E_{ab} &=& - E_{ab} \, \Theta + \curl H_{ab} - \frac{1}{2} \l( \rho + p \r) \, \sigma_{ab} - \frac{1}{6} \Theta \, \pi_{ab} - \frac{1}{2} \dot \pi_{ab} \ , \label{tpeq4}\\
\dot H_{ab} &=& - H_{ab}\,\Theta - \curl E_{ab} + \frac{1}{2} \curl \pi_{ab} \ . \label{tpeq5}
\end{eqnarray}
Equations \eqref{tceq5} and \eqref{tpeq3} determine the magnetic and electric Weyl tensors, respectively, from the shear.  We now have a closed system of equations describing the evolution of the gravitational waves.
%

\subsection{Gravitational Waves}
Upon the differentiation of equations \eqref{tpeq3}--\eqref{tpeq5} with respect to the cosmic time, using the energy conservation equation (Eq. \ref{tpeq1}), the Raychaudhuri equation (Eq. \ref{tpeq2}), and the commutator identities \cite{Challinor:1999xz,Maartens:1997sh,VanElst:1998}:
\begin{equation}
\curl \dot X_{ab} = (\curl X_{ab})\dot{} + \frac{1}{3} \Theta \curl X_{ab} \ ,
\end{equation}
\begin{equation}
\curl \curl S_{ab} = \D^2 S_{ab} - \frac{3}{2} \D_{\la a} \D^c S_{b\ra c} \ ,
\label{id:5}
\end{equation}
one arrives to gravitational wave equations for the shear, the electric and magnetic Weyl tensors in flat models:
\begin{align}
\ddot \sigma_{ab} &- \D^2 \sigma_{ab}  + \frac{5}{3} \, \Theta \, \dot \sigma_{ab} + \l( \frac{1}{2} \rho - \frac{3}{2} p 
\r) \, \sigma_{ab} = \dot \pi_{ab} + \frac{2}{3} \, \Theta \, \pi_{ab} \ , \label{gws}\\
\ddot E_{ab} &- \D^2 E_{ab} + \frac{7}{3} \, \Theta \,\dot E_{ab} + \frac{2}{3} \, \l( \rho - p \r) E_{ab}
- \frac{1}{6} \Theta \, \l( \rho + p \r) \, \l( 1 + 3 \, c_s^2 \r) \, \sigma_{ab} \nn\\
&= -\l[ \frac{1}{2} \ddot \pi_{ab} -\frac{1}{2} \D^{2} \pi_{ab} + \frac{5}{6} \, \Theta \, \dot \pi_{ab} + \frac{2}{3}\rho \,\pi_{ab}\r] \ , \label{gwe}\\
\ddot H_{ab} &- \D^2 H_{ab} + \frac{7}{3} \, \Theta \, \dot H_{ab} + 2 \l(\rho - p \r) \, H_{ab}
= \curl \dot \pi_{ab} + \frac{2 }{3} \, \Theta \, \curl \pi_{ab} \ , \label{gwh}
\end{align}
where $c_{s}^{2}=\dot{p}/\dot{\rho}$ and $\D^2 \equiv \D^a \D_a$.  

It can be seen from equations \eqref{gwe} and \eqref{gwh} that the electric and magnetic Weyl tensors play an important role in the propagation of gravitational waves. The curl terms in the corresponding propagation equations give rise to the wave behaviour just like they do in electromagnetism \cite{Dunsby:1998hd}. equation \eqref{gwe} is not closed due to the shear term hence the necessity of deriving the shear wave equation (Eq. \ref{gws}) as well. Moreover, it is more convenient solving for the shear then getting $H_{ab}$ via \eqref{tceq5} rather than starting from equation \eqref{gwh}.

\subsection{Mode Expansion in Tensor Harmonics}
Spatial harmonics are eigenfunctions of the Laplace-Beltrami operator:
\begin{equation}
\D^2 Q = - \frac{k^2}{a^2} Q \ ,
\label{LBeq}
\end{equation}
and are, by construction, covariantly constant:
\begin{equation}
\dot Q = 0 \ .
\end{equation}
$k = \frac{2 \pi a}{\lambda}$ is the wave number and $Q$ represents scalar $\l( Q^{(0)} \r)$, vector $\l( Q^{(1)}_a \r)$, or tensor $\l( Q^{(2)}_{ab} \r)$ harmonics \cite{Bruni:1992dg,Harrison:1967zz}.

Because different modes are independent, first order quantities $X(t,x)$ can be decomposed to or reconstructed from its harmonics via:
\begin{equation}
    X(t,x) = \sum_k X^{(k)}(t) Q^{(k)}(x) \ ,
\end{equation}
assuming that the quantity $X$ is factorisable into purely temporal and spatial components \cite{Carloni:2007yv}.

That said, transverse PSTF quantities such as $E_{ab}$, $H_{ab}$, $\sigma_{ab}$, and $\pi_{ab}$ can be expanded in terms of electric and magnetic parity tensor harmonics; $Q_{ab}$ and $\bar Q_{ab}$ respectively:
\begin{subequations}
\begin{align}
E_{ab} &= \sum_{k} \l( \frac{k}{a} \r)^2 \l[ E_k Q_{ab}^{(k)} + \bar E_k \bar Q_{ab}^{(k)} \r] \ , \\ 
H_{ab} &= \sum_{k} \l( \frac{k}{a} \r)^2 \l[ H_k Q_{ab}^{(k)} + \bar H_k \bar Q_{ab}^{(k)} \r] \ , \\ 
\sigma_{ab} &= \sum_{k} \frac{k}{a} \l[ \sigma_k Q_{ab}^{(k)} + \bar \sigma_k \bar Q_{ab}^{(k)} \r] \ , \label{shear_harmonics} \\
\pi_{ab} &= \rho \sum_{k} \l[ \pi_k Q_{ab}^{(k)} + \bar \pi_k \bar Q_{ab}^{(k)} \r] \ ,
\end{align}
\label{tensor_harmonics_expansion}
\end{subequations}
where $k$ denotes distinct harmonic modes. Tensor harmonic expansion conveniently converts the propagation equations into ordinary differential equations and the constraint equations into algebraic relations. We have adopted the same expansion as \cite{Challinor:1999xz,Leong:2002hs}.

The electric and magnetic parity tensor harmonics are related via:
\begin{subequations}
\begin{eqnarray}
\curl Q^{(k)}_{ab} &=& \frac{k}{a} \, \bar Q^{(k)}_{ab} \ , \\
\curl \bar Q^{(k)}_{ab} &=& \frac{k}{a} \, {Q}^{(k)}_{ab} \ .
\end{eqnarray}
\end{subequations}
Substituting into equation \eqref{tceq5} gives:
\begin{subequations}
\label{tceq5k}
\begin{eqnarray}
& H_k = \bar \sigma_k \label{tceq5k1}  \ , \\ 
& \bar H_k = \sigma_k \label{tceq5k2}  \ .   
\end{eqnarray}
\end{subequations}

Expanding the shear wave equation (Eq. \ref{gws}) into tensor harmonics using equation \eqref{shear_harmonics} and keeping in mind equation \eqref{LBeq}, and the Friedmann equations \eqref{friedmann_eqs}, we get:
\begin{equation}
\ddot \sigma_k + \Theta \dot \sigma_k + \l[ \frac{k^2}{a^2} - \frac{1}{3} \l( \rho + 3 p \r) \r] \sigma_k = \frac{a}{k} \l[ \rho \dot \pi_k - \frac{1}{3} \l( \rho + 3 p \r) \Theta \pi_k \r] \ .
\label{gwsk}
\end{equation}
This equation is similar in form to the one in \cite{Challinor:1999xz}, up to a minus sign due to the use of opposite metric tensor signatures. It is of great importance since any solution for the shear would yield the magnetic and electric Weyl tensors according to equations \eqref{tceq5} and \eqref{tpeq3}. It is worth mentioning again that $E_{ab}$ and $H_{ab}$, together with the Ricci tensor provide a full description of the curvature of spacetime as stated by equation \eqref{Riemann_decomp}. 

The wave equation for the shear (Eq. \ref{gwsk}) is general for flat FLRW regardless of the fluid. We will see in the next chapter that in the case of \fR gravity, the density and the pressure terms in the left hand side and the anisotropic stress term in the right hand side of equation \eqref{gwsk} can be extended to include modifications of gravity via their contribution to the energy momentum tensor. 



\chapter{\fR Theories} 
\label{fR}
%

\ifpdf
    \graphicspath{{4/figures/PNG/}{4/figures/PDF/}{4/figures/}}
\else
    \graphicspath{{4/figures/EPS/}{4/figures/}}
\fi

%
%
\section{Overview of \fR Gravity}

\fR theories of gravity, formally proposed in \cite{Buchdahl:1983zz}, are a class of ETGs that represents one of the simplest modifications to GR. They arise from replacing the Ricci scalar $R$ in the Einstein-Hilbert Lagrangian with a more general and arbitrary function \fR  \cite{Sotiriou:2008rp,DeFelice:2010aj}. While different formalisms of the theory of general relativity lead to the same field equations due to the linearity of the Lagrangian in $R$, one would not expect the same for higher order theories of gravity \cite{Capozziello:2009nq}. Indeed, there are three flavours of \fR gravity. These are the metric, Palatini, and metric affine formalisms concisely presented below. For a more extensive description, the reader is referred to \cite{Sotiriou:2007yd}. 

\subsection{The Metric Formalism}
\label{fR_metric_formalism}
Variation with respect to the metric of the \fR action:
\begin{equation}
    \label{actionfR} 
    S_{metric} = \int_\mathcal{V} \l[ f(R) + \mathcal{L}_m(g_{ab},\Psi_m) \r] \sqrt{-g} \ \text d^4x \ ,
\end{equation}
leads to fourth order (partial differential) gravitational field equations \cinc \footnote[\countval]{See Appendix \ref{FE_derivation}.}:
\begin{equation}
R_{ab} - \frac{1}{2} \ g_{ab} \ R  = \frac{1}{f_R} \ T_{ab} + T^R_{ab} \ ,
\label{FE}
\end{equation}
where $f_R \equiv \frac{df}{dR}$ and:
\begin{equation}
T^R_{ab} \equiv \frac{1}{f_R} \l[ \frac{1}{2} \ g_{ab} \ f + \l(\nabla_a \nabla_b - g_{ab} \Box \r) f_R  - \frac{1}{2} \ f_R \ g_{ab} \ R \r] \ .
\end{equation}

The surface term that appears from the action (Eq. \ref{actionfR}) is not a total variation as it is the case for GR and therefore cannot be ``healed'' by adding a total divergence such as the GHY surface term. Since the action contains higher order derivatives of the metric, the problem can be overcome by fixing more degrees of freedom on the boundary \cite{Sotiriou:2008rp}. It can easily be verified that the field equations \eqref{FE} reduces to the EFE for $f(R)=R$.

\subsection{The Palatini Formalism}
As it is the case for GR, the Palatini formalism of \fR gravity considers the connection as an additional structure (on the manifold) independent of the metric. The \fR Palatini action takes the form:
\begin{equation}
S_{Palatini} = \int_\mathcal{V}  \l[ f(\mathcal{R}) +  \mathcal{L}_m(g_{ab},\Psi_m) \r] \sqrt{-g} \text d^4x \ ,
\label{actionPalatini} 
\end{equation}
where, again, $\mathcal{R}$ is the Ricci scalar defined in terms of the connection. Variation of the previous action with respect to the connection leads to:
\begin{equation}
\bar \nabla_c \l( \sqrt{-g}  g^{ab} f_\calR \r) = 0 \ ,
\label{PalatiniconnectionFE} 
\end{equation}
where $f_\calR \equiv \frac{df}{d\calR}$. The bar over the del operator denotes the covariant derivative defined only in terms of the connection. On the other hand, variation with respect to the metric gives the field equations:
\begin{equation}
f_\calR \calR_{ab} - \frac{1}{2} f(\calR) g_{ab} = T_{ab} \ .
\label{PalatinimetricFE} 
\end{equation}

For $f(\calR) = \calR$, Eq. \eqref{PalatiniconnectionFE} imposes the metric postulate \eqref{metric_postulate} and that the Christoffel symbol is indeed the Levi-Civita connection. Then Eq. \eqref{PalatinimetricFE} becomes the EFE \cite{Capozziello:2009nq,Clifton:2011jh}.

\subsection{The Metric Affine Formalism}
For the metric affine action,
\begin{equation}
S_{m.affine} = \int_\mathcal{V} \l[ f(\mathcal{R}) + \mathcal{L}_m(g_{ab},\Gamma,\Psi_m) \r] \sqrt{-g} \text d^4x \ ,
\label{actionMA} 
\end{equation}
and like GR, the field equations are prone to inconsistencies arising from the fact that matter fields are not, generally, invariant under projective transformations \eqref{projective_transform} while the Ricci scalar is invariant. This is dealt with by adding a Lagrange multiplier to the action \cite{Sotiriou:2006qn,Clifton:2011jh}:
\begin{equation}
S_{LM} = \int_\mathcal{V} \sqrt{-g} B^a S_a \text d^4x \ ,
\label{lagrange_multiplier}
\end{equation}
where $S_a=S_{ba}{}^b$. Varying the total action with respect to the Lagrange multiplier $B^b$, the metric, and the connection, respectively, leads to the field equations:
\begin{equation}
S_{ba}{}^a = 0 \ ,
\label{MALMFE}
\end{equation}
\begin{equation}
f_\calR \calR_{ab} - \frac{1}{2} f(\calR) g_{ab} = T_{ab} \ ,
\label{MAmetricFE} 
\end{equation}
\begin{equation}
\delta^b_{\phantom bc} \l( \sqrt{-g} f_\calR g^{ad} \r)_{;d} - \l( \sqrt{-g} f_\calR g^{ab} \r)_{;c} + 2 \sqrt{-g} f_\calR g^{ad} \Gamma^b_{\phantom b[dc]} = \sqrt{-g} \l( \Delta_c^{\phantom cab} - B^{[b} \, \delta^{a]}_{\phantom{a]}c} \r) \ ,
\label{MAconnectionFE} 
\end{equation}
where $\Delta_c^{\phantom cab} \equiv \frac{-2}{\sqrt{-g}} \frac{\delta \mathcal{L}_m}{\delta \Gamma^c_{\phantom cab}}$ is called the \emph{hypermomentum}  \cite{Vitagliano:2010sr}. Taking the trace of the last equation and using Eq. \eqref{MALMFE}, the Lagrange multiplier turns out to be:
$$B^b = \frac{2}{3} \Delta_d^{\phantom ddb} \ .$$
Thus the field equations \eqref{MAconnectionFE} become:
\begin{equation}
\delta^b_{\phantom bc} \l( \sqrt{-g} f_\calR g^{ad} \r)_{;d} - \l( \sqrt{-g} f_\calR g^{ab} \r)_{;c} + 2 \sqrt{-g} f_\calR g^{ad} \Gamma^b_{\phantom b[dc]} = \sqrt{-g} \l( \Delta_c^{\phantom cab} -\frac{2}{3} \Delta_d^{\phantom dd[b} \, \delta^{a]}_{\phantom{a]}c} \r) \ .
\label{MAconnectionFEfinal} 
\end{equation}

The metric affine formalism is more general than the metric and Palatini ones and reduces to these if further assumptions about the matter Lagrangian density and the connection are made. But it is also more complicated and has not been investigated as much.

In this work, we are interested in cosmological perturbations in metric \fR theories of gravity. For the remainder of this text, plain \fR refers to metric $f(R)$ as presented in \ref{fR_metric_formalism}.
\section{The Dynamics of Metric $f(R)$ Cosmologies}
The field equations \eqref{FE} are written in such a way to keep the traditional Einstein tensor in the left hand side. The first term in the right hand side is the classical EMT divided by the first derivative of \fR and thus may be interpreted as the effective contribution of standard matter in the new geometry. The second term is fully geometrical and will be referred, hereafter, as the \emph{curvature fluid}. This multi-fluid decomposition is useful in the sense that the covariant approach can be applied the usual way once the thermodynamics of the fluids are determined \cite{Ananda:2007xh}.

Applying the operations in equations \eqref{EMT_decomp_ops} to the total EMT on the right hand side of Eq. \eqref{FE}, We get:
\begin{subequations}
\label{EMT_decomp}
\begin{align}
\rho &= \frac{\rho^m}{f_R} + \rho^R \ , \label{fRdensity} \\
p &= \frac{p^m}{f_R} + p^R \ , \\
q_a &= \frac{q_{a}^m}{f_R} + q_{a}^R \ , \\
\pi_{ab} &= \frac{\pi_{ab}^m}{f_R} + \pi_{ab}^R \ , \label{anis_decomp}
\end{align}
\end{subequations}
where the superscript ``$R$'' indicates the contribution of the curvature fluid to different components of the EMT. The linearised components of the curvature fluid are found to be  \cite{Ananda:2007xh}: 
\begin{subequations}
\label{curvature_fluid}
\begin{align}
\rho^{R} &= \frac{1}{f_R} \l[ \frac{1}{2}(R f_R-f)-\Theta f_{RR}\dot{R}+f_{RR}\tilde{\nabla}^2{R} \r] \ , \\
p^{R} &= \frac{1}{f_R} \l[ \frac{1}{2}(f-Rf_R)+f_{RR}\ddot{R}+f_{RRR}\dot{R}^2+\frac{2}{3}\Theta f_{RR}\dot{R} -\frac{2}{3}f_{RR}\tilde{\nabla}^2{R} \r] \ , \\
q^{R}_a &= \frac{1}{f_R} \l[ - f_{RRR}\dot{R}\D_{a}R - f_{RR}\D_{a}\dot{R} + \frac{1}{3}f_{RR} \D_{a}R \r] \ , \\
\pi^{R}_{ab} &= \frac{1}{f_R} \l[ f_{RR}\D_{\lgl a}\D_{b\rgl}R - f_{RR} \, \sigma_{ab}\dot{R} \r] \ , \label{curvature_anis}
\end{align}
\end{subequations}
all vanishing in GR.

\section{The Cosmological Field Equations in \fR  Gravity}
The generalised Friedmann equations in $f(R)$, maintain the same form as in equations \eqref{friedmann_eqs}. But the components of the EMT now have curvature fluid terms. Separating these from the standard matter \cite{delaCruzDombriz:2010xy}, they read:

\begin{subequations}
\label{friedmann_eqs_tau}
\begin{align}
& \frac{1}{2} f - 3 f_R \frac{\ddot a}{a} + 3 \frac{\dot a}{a} \dot{R} f_{RR} = \rho_m \label{friedmann_eqs_velocity_fR} \ , \\
& \frac{1}{2} f - f_R (\dot{H}+3H^{2}) + \frac{1}{a} \frac{\text d}{\text dt} \l(a^2 \dot{R} f_{RR}\r) = - p_m \ , \label{friedmann_eq_acceleration_fR}
\end{align}
\end{subequations}
for flat FLRW with vanishing cosmological constant.

\subsection*{The Background Evolution in $R^n$ Gravity}
We extend the discussion to the case of $f(R) = R^n$ which we will later consider in the numerical simulations. Eq. \eqref{friedmann_eqs_velocity_fR} becomes:
\begin{equation}
\frac{1}{2} R^n - 3 n \frac{\ddot a}{a} R^{n-1} + 3 n (n-1) \frac{\dot a}{a} \dot R R^{n-2} = \rho_m = \rho_{dust} + \rho_\gamma \ ,
\label{friedmann_Rn}
\end{equation}
where we have assumed that the standard matter is composed only of dust and radiation.

In addition, the Ricci scalar in FLRW is related to the scale factor via:
\begin{equation}
R = 6 \l[ \l( \frac{\dot a}{a} \r)^2 + \frac{\ddot a}{a} \r]
\label{Ricci-a-cosmic}
\end{equation}
\cite{Capozziello:2007ec}. Substituting for $R$ in equation \eqref{friedmann_Rn} and using the density evolution equation \eqref{density_evolution} for individual conserved energy species, we get the equation for the evolution of the scale factor in $R^n$ gravity:
\begin{align}
\l[ 6 \l( \l( \frac{\dot a}{a} \r)^2 + \frac{\ddot a}{a} \r) \r]^{n-2} & \l[ n(1-n) \l( a^2 \dot a \dddot a \r) + (2n^2 - 2n -1) \dot a^4 + (n-1) a^2 \ddot a^2 \r. \nn \\
& \l. + (2n - 2 - n^2) a \dot a^2 \ddot a \r] = -\frac{1}{18} \l[ \rho_{\gamma,0} + \rho_{dust,0} a \r] \ ,
\label{background}
\end{align}
or in conformal time:
\begin{equation}
\l( 6 \frac{a''}{a^3} \r)^n \l[ (1-n) + \frac{(a')^2}{a a''} ( 4n - 3n^2 ) + n(n-1) \frac{a' a'''}{(a'')^2} \r] = 2 \rho_m \ .
\label{background_conformal}
\end{equation}
Both equations \eqref{background} and \eqref{background_conformal} reduce to the velocity Friedmann equation (Eq. \ref{friedmann_dot}) for $n=1$.

The non trivial solution of Eq. \eqref{background_conformal} requires $R \neq 0$ (non-empty universe). Then taking non GR solutions ($n \neq 1$), the last equation becomes, with further simplification:
\begin{equation}
a''' + \l( \frac{4-3n}{n-1} \r) \frac{a' a''}{a} + \frac{1}{n} \frac{(a'')^2}{a'} - \l( \frac{1}{18 n (n-1)} \r) \frac{a^2}{a'} \l( \frac{a^3}{6 a''} \r)^{n-2} (\rho_{\gamma,0} + \rho_{dust,0} a) = 0 \ .
\end{equation}
Numerical solution of Eq. \eqref{background_conformal} would provide values $a(\eta)$ describing the evolution of the scale factor with respect to the conformal time $\eta$.

\section{Viability of $f(R)$ Gravity}
\subsection{Viability Conditions of $f(R)$ Gravity}
\label{viability_conds}
\fR theories are required to satisfy some conditions in order to be consistent with existing data. \cite{Pogosian:2007sw} lists the following constraints for cosmic acceleration \fR models.
\begin{enumerate}
\item $f_{RR} > 0$ for $R \gg f_{RR}$ must be satisfied in order to have a stable high curvature regime, such as the matter dominated era, in the cosmological evolution. 
\item $f_R > 0$ has to be true for the effective Newtons constant, $G_{eff} = G/f_R$, to be positive at all times. This can be seen from the field equations \eqref{FE}. Violation of this condition results in the universe quickly becoming inhomogeneous and anisotropic.  
\item $f_R \leq 1$, together with the two previous conditions, constrain $f_R$ to be monotonically increasing, approaching $1$, as $R \rightarrow \infty$. This GR behaviour at early times is dictated by the big bang nucleo-synthesis and the CMB data.
\item $|f_R -1|\ll 1$ to have late time acceleration and satisfy local gravitational constraints.
\end{enumerate}
These constraints take a form slightly different of those in \cite{Pogosian:2007sw} as the gravitational Lagrangian therein is $R + f(R)$.
 
\subsection{Some Viable \fR Models}
Here are presented some viable \fR models from the literature.

\subsubsection{$f(R)  = \frac{1}{2} \l( R - m^2 \frac{c_1  (R/m^2)^n}{c_2 (R/m^2)^n +1} \r)$}
\cite{Hu:2007nk} presented an \fR model in which the universe accelerates without resorting to a cosmological constant. The action takes the form
\begin{equation}
S = \int_\mathcal{V} \l[ \frac{R + F(R)}{2} \r] \sqrt{-g} \ \text d^4x + S_m \ ,
\end{equation}
where $F(R)$ has to be chosen in such a way that the model mimics $\Lambda$CDM at high redshifts, that is, to be in agreement with the CMB data. This is imposed by:
\begin{equation}
\lim_{R \to \infty} F(R) = constant \ ,
\end{equation}
which ensures standard GR behaviour at high curvature regimes. In addition, the model should be consistent with local gravitational tests:
\begin{equation}
\lim_{R \to 0} F(R) = 0 \ .
\end{equation}

The above conditions can be satisfied, according to \cite{Hu:2007nk}, by a general class of broken power law \fR models:
\begin{equation}
F(R)  = - m^2 \frac{c_1  (R/m^2)^n}{c_2 (R/m^2)^n +1} \ ,
\label{hu_sawicki}
\end{equation}
with $n>0$ and $ m^2 \equiv \rho_0/3$. $c_1$ and $c_2$ are dimensionless parameters. The sign of $F(R)$ is chosen so that its second derivative is positive in compatibility with the first of the viability conditions presented in subsection \ref{viability_conds}; which makes the model stable at high curvature: $R \gg m^2$.

Although no cosmological constant was explicitly introduced in this model, the expansion of the right hand side of equation \eqref{hu_sawicki} at high curvature regimes:
\begin{equation}
\lim_{\frac{m^2}{R} \to 0} F(R) \approx -\frac{c_1}{c_2} m^2 + \frac{c_1}{c_2} m^2 \l( \frac{m^2}{R} \r)^n \ ,
\end{equation}
shows that $F(R)$ is the cosmological constant in the limiting case of $c_1/c_2^2 \rightarrow 0$ for fixed $c_1/c_2$. On the other hand, a finite $c_1/c_2^2$ prevents the curvature from declining all the way with the matter density. These models encompass an accelerating universe, similarly to $\Lambda$CDM.

\subsubsection{$f(R) = R + \lambda R_0\left[\left(1+\frac{R^2}{R_0^2}\right)^{-n}-1\right]$}
Consider the Lagrangian proposed by \cite{Starobinsky:2007hu}:
\begin{equation}
f(R) = R + \lambda R_0\left[\left(1+\frac{R^2}{R_0^2}\right)^{-n}-1\right] \ ,
\label{starobinsky}
\end{equation}
with $n$ and $\lambda$ strictly positive and $R_0$ is of order comparable to the cosmological constant. 

Flat and empty spacetime ($R = 0$ and $T_{ab} = 0)$ admits $R_{ab} = 0$ as a solution to the FE \eqref{FE} but is unstable since $f_{RR} \leq 0$. In high curvature regimes ($R>>R_0$), equation \eqref{starobinsky} becomes:
\begin{equation}
f(R) = R - \lambda R_0 \ ,
\label{starobinsky_high_R}
\end{equation}
where $\lambda R_0$ may be regarded as the equivalent of twice the cosmological constant at high curvatures. 

The de Sitter solutions for the parameter $\lambda$ are of the form:
\begin{equation}
\lambda=\frac{x_1(1+x_1^2)^{n+1}}{2\left((1+x_1^2)^{n+1}-1-(n+1)x_1^2\right)}~,
\label{lambdadS}
\end{equation}
with $x_1 = R/R_0 = const > 0$. It is easy to see from equation \eqref{lambdadS} that $x_1<2\lambda$. The effective cosmological constant at $R/R_0$ is one quarter of the Ricci scalar, i.e. $\Lambda=R/4$. Asymptotically, $x_1 \to 2\lambda$ and the evolution exhibits $\Lambda$CDM behaviour.

\section{Reconstruction of \fR Models}

\subsection{\fR Models Mimicking $\Lambda$CDM}

Reconstruction of an \fR theory that admits a $\Lambda$CDM model may be done, as presented in \cite{delaCruzDombriz:2006fj,Dunsby:2010wg}, by starting from the observation supported relation for the Hubble parameter:
\begin{equation}
H(z)=\sqrt{\frac{\rho_0}{3}(1+z)^3 + \frac{\Lambda}{3}} \ .
\end{equation}
The first and second time derivatives of the scale factor are then:
\begin{align}
\dot a & = \sqrt{\frac{\rho_0}{3a}+\frac{\Lambda}{3}} \ ,\\
\ddot a & = \frac{2\Lambda a^3-\rho_0}{6a^2} =\frac{1}{2} \frac{\text d \dot a^2}{\text d a} \ .
\end{align}
Substituting for $\dot a$ and $\ddot a$ in equation \eqref{Ricci-a-cosmic}, $R(a)$ and equivalently $a(R)$ are obtained: 
\begin{equation}
R(a)=\frac{4\Lambda a^3+\rho_0}{a^3} \quad \Longleftrightarrow \quad a(R)=\left(\frac{\rho_0}{R-4\Lambda}\right)^{(1/3)} \ .
\label{RaaR}
\end{equation}

Combining the above equations and plugging in the \fR Friedmann equation \eqref{friedmann_eqs_velocity_fR} in the presence of a positive cosmological constant, one gets the equation:
\begin{equation}
-3(R - 3\Lambda)(R-4\Lambda)f_{RR} + \l(\frac{R}{2}-3\Lambda\r) f_R + \frac{1}{2} f - \rho (R) = 0 \ ,
\label{friedmann_eqs_R}
\end{equation}
admitting an \fR solution in the form:
\begin{equation}
f(x) = C_1 F\l([\alpha_+,\alpha_-],-\frac{1}{2};x\r) + C_2x^{\frac{3}{2}} F\l([\beta_+,\beta_-],\frac{5}{2};x\r) \ ,
\end{equation}
where $F$ is the hypergeometric function of variable $x\equiv -3 + R/\Lambda$ with parameters $\alpha_{\pm}=(-7\pm\sqrt{73})/12$, $\beta_{\pm}=(-11\pm\sqrt{73})/12$, and arbitrary integration constants $C_{1}$ and $C_{1}$.

Keeping in mind that the scale factor in equations \eqref{RaaR} is positive, $F$ can either be complex or divergent. To ensure a real valued function \fR demands that the integration constants vanish. Therefore, there exists no real \fR function that mimics $\Lambda$CDM.


\subsection{\fR Models Mimicking a GR Universe Containing a Single Perfect Fluid}
Here, we shall only present some examples of reconstructions of \fR models that emulate the evolution of a GR universe containing a perfect fluid with a known equation of state. For a more general study, the reader is referred to \cite{delaCruzDombriz:2010xy,Dunsby:2010wg}.
\subsubsection*{Dust}
In a dust filled universe with a cosmological constant, it follows from equations \eqref{RaaR} and \eqref{density_evolution}, for $\omega=0$, that:
\begin{equation}
\rho_m(R) = R - 4\Lambda \ .
\end{equation}
Substituting into \eqref{friedmann_eqs_R}, one gets the GR Lagrangian in the presence of a cosmological constant (Eq \ref{GLDE}). If, however, the cosmological constant is set to zero, the general solution is of the form:
\begin{equation}
f(R)=R+C_1R^{\alpha_+}+C_2R^{\alpha_-} \ ,
\end{equation}
proving that there exists some real valued functions \fR able to mimic a dust filled universe \cite{Dunsby:2010wg}

\subsubsection*{A Fluid with $\omega = -\frac{1}{3}$}
Similarly for $\omega=-1/3$, equations \eqref{RaaR} and \eqref{density_evolution} imply:
\begin{equation}
\rho_\gamma(R) = [\rho_{\gamma,0}(R-4\Lambda)]^{2/3} \ .
\end{equation}
Substituting into \eqref{friedmann_eqs_R} yields:
\begin{equation}
f(R)=\mu(R-4\Lambda)^{2/3} \ ,
\end{equation}
where $\mu$ is a constant that depends on $\rho_{\gamma,0}$.



\chapter{CMB Tensor Anisotropies in \fR Gravity} 
\label{anis_fR}


\ifpdf
    \graphicspath{{5/figures/PNG/}{5/figures/PDF/}{5/figures/}}
\else
    \graphicspath{{5/figures/EPS/}{5/figures/}}
\fi

\section{Motivation}
As it was seen before, perturbations may be decomposed into independently evolving scalar, vector, and tensor modes. Vector perturbations decay in an expending universe and hence are not important in the evolution of the total perturbations. Tensor and scalar perturbations in GR were presented in \cite{Lewis:2000PhD}. Scalar perturbations in \fR gravity were studied in \cite{Abebe:2011ry,delaCruzDombriz:2008cp}. 
In this work, we extend the previous work done on CMB tensor anisotropies in general relativity \cite{Lewis:2000PhD} to the case of \fR gravity.

\section{Evolution}
For pure tensor modes, equation \eqref{curvature_anis} reduces to a crucial equation, for this work, that relates the tensor part of the curvature anisotropy and the shear:
\begin{equation}
\pi^{R}_{ab} = - \frac{f_{RR}}{f_R} \, \sigma_{ab}\dot{R}  \ .
\label{curvature_anis_tensor}
\end{equation}
Decomposed into tensor harmonics, it becomes:
\begin{equation}
\pi^{R}_k = - \frac{k}{a} \frac{1}{\rho} \frac{f_{RR}}{f_R} \dot{R} \, \sigma_k \ ,
\label{curvature_anis_tensor_k}
\end{equation}
providing the missing piece for the numerical solution of equation \eqref{gwsk}.

\section{Initial Conditions}

In the radiation dominated era: $\pi^m=\pi^\gamma$. Using equations \eqref{anis_decomp} and \eqref{curvature_anis}, equation \eqref{gwsk} becomes, to first order:
\begin{equation}
\ddot \sigma_k + A \dot \sigma_k + B \sigma_k = \frac{a}{k} \frac{\rho}{f_R} \l[ \dot \pi^\gamma_k - \l( H + 3H\omega + \frac{f_{RR}}{f_R} \r) \pi^\gamma_k \r]  \ ,
\label{shear_wave_eq_k}
\end{equation}
with
\begin{equation}
A = 3H + \frac{f_{RR}}{f_R} \dot R \ ,
\end{equation}
and
\begin{equation}
B = \frac{k^2}{a^2} + 2 \frac{\ddot a}{a} + \dot R^2 \l[ \frac{f_{RRR}}{f_R} -\l( \frac{f_{RR}}{f_R} \r) ^2 \r] + \frac{f_{RR}}{f_R} \ddot R + H \frac{f_{RR}}{f_R} \dot R  \ .
\end{equation}

However, the early universe was sufficiently homogeneous and isotropic that the radiation anisotropic stress, $\pi^\gamma$, may be safely  neglected \cite{Challinor:1999xz,Leong:2002hs}. Changing to conformal time, Eq. \eqref{shear_wave_eq_k} becomes:
\begin{equation}
\sigma_k '' + \l( a A - \mathcal{H} \r) \sigma_k ' + a^2 B \sigma_k =  0 \ ,
\label{shear_wave_eq_eta}
\end{equation}
where $\mathcal{H} \equiv \frac{a'}{a} = aH$. Making the variable change:
\begin{equation}
u_k = a^m \sigma_k \ ,
\label{sigma2u}
\end{equation}
and choosing:
\begin{equation}
m = \frac{Aa - \mathcal{H}}{2 \mathcal{H}} \ ,
\label{sigma_power}
\end{equation}
to get rid of the damping term, we get:

\begin{equation}
u_k '' + \l[ - \frac{1}{2} m \mathcal{H} \frac{f_{RR}}{f_R} R' - m \frac{a''}{a} + a^2 B \r] u_k = 0 \ .
\label{uwave}
\end{equation}

\section{The Case of $f(R) = R^n$}

So far, we have generally considered \fR theories in flat geometries. Proceeding further requires the choice of a model. For the sake of simplicity, we take $f(R)$ to be a power law of the Ricci scalar; i.e. $f(R) = R^n$. This is, indeed, the simplest extension to general relativity in the \fR class of gravity theories \cite{Clifton:2005aj,Barrow:2005dn}; and reduces to GR when $n=1$.

\subsection{The Evolution of Perturbations in $R^n$ Gravity}
Substituting for \fR in equation \eqref{curvature_anis_tensor_k} gives:
\begin{equation}
\pi^{R}_k = - \frac{k}{a^2} \l( \frac{n-1}{\rho} \r) \frac{R'}{R}  \, \sigma_k \ .
\label{curvature_anis_tensor_k_Rn}
\end{equation}
In addition, combining the conformal time equivalent of equation \eqref{Ricci-a-cosmic}:
\begin{equation}
 R = 6 \frac{a''}{a^3} \ ,
\label{Ricci-a}
\end{equation}
with the Friedmann equations for flat models with no cosmological constant, expressed in conformal time:
\begin{subequations}
\label{friedmann_eqs_tau}
\begin{align} 
& \mathcal{H}^2 = \frac{1}{3} \rho a^2 \ , \\
& \mathcal{H} ' = - \frac{a^2}{6} \l(\rho + 3 p \r) \ , 
\end{align}
\end{subequations}
we get the Ricci scalar in terms of the total density and pressure:
\begin{equation}
R = \rho - 3p \ .
\label{R_EMT_trace}
\end{equation}

As a side note, it can be seen from the last equation that radiation, having equation of state of $\omega_\gamma = \frac{1}{3}$, does not contribute to the Ricci scalar:
\begin{equation}
\rho_\gamma - 3p_\gamma = 0 \ .
\label{R_rad}
\end{equation}
So the curvature fluid is the only source of curvature in the radiation dominated era. At later times, for a fluid composed only of dust and radiation, and keeping in mind that dust is pressure-less, i.e. $\omega_{dust} = 0$,  equation \eqref{R_EMT_trace} becomes:
\begin{equation}
\rho_{dust} = f_R [R  - (\rho^R - 3 p^R) ] \ ,
\end{equation}
providing a relation between the dust density and geometry.

Now, remembering equation \eqref{EoS}, we have:
\begin{equation}
R = \rho (1-3\omega) \ ,
\end{equation}
and thus:
\begin{equation}
R' = \rho' - 3\omega' \rho - 3 \omega \rho'  \ .
\end{equation}
Using the conformal time version of the total energy conservation equation \eqref{Econserv}, we get:
\begin{equation}
\frac{R}{R'} = \l[ -3 \l(\mathcal{H} (1+\omega) + \frac{\omega'}{1-3\omega} \r) \r]^{-1} \ .
\end{equation}
Substituting into \eqref{curvature_anis_tensor_k_Rn}, the evolution of the curvature anisotropy is now:
\begin{equation}
\pi^{R}_k = \l[ 3 \l(\mathcal{H} (1+\omega) + \frac{\omega'}{1-3\omega} \r) \r]^{-1} \, \frac{k}{a^2} \l( \frac{n-1}{\rho} \r)  \, \sigma_k \ .
\label{anis_Rn_rho}
\end{equation}

For the last equation to be useful, one needs an expression for the effective EoS parameter of the fluid in terms of the conformal time $\eta$ or a function of time such as $a(\eta)$.  The effective EoS parameter of a fluid containing different species $i$ is:
\begin{align}
\omega_{eff} &= \frac{p^{tot}}{\rho^{tot}} \label{EoSp_general} \\
& = \frac{\displaystyle \sum_i \omega_i \rho_i}{\displaystyle \sum_i \rho_i}  \label{EoSp_expansion} \\
& = \frac{\displaystyle \sum_i \omega_i \rho_{i,0} a^{-3(1+\omega_i)}}{\displaystyle \sum_i \rho_{i,0} a^{-3(1+\omega_i)}} \label{EoSp_density_today}\\
& = \frac{\displaystyle \sum_i \omega_i \Omega_{i,0} a^{-3(1+\omega_i)}}{\displaystyle \sum_i \Omega_{i,0} a^{-3(1+\omega_i)}} \ . \label{EoSp_density_param}
\end{align}
where equations \eqref{density_evolution} and \eqref{density_params} have been used to obtain these relations. $\Omega_{i,0}$ denotes the present density parameter of energy species $i$.

The expansion in equation \eqref{EoSp_expansion} can be performed only if the densities of the different species are not interacting (even though the fluids themselves might be). Take for example the total density in \fR gravity (Eq. \ref{fRdensity}). The standard matter density interacts with the curvature fluid through the term $1/f_R$. In such cases, the EoS parameter is obtained directly at the level of equation \eqref{EoSp_general}. 

Moreover, the expression in equation \eqref{EoSp_density_today} is only valid for conserved species. Although the RHS of \fR field equations \eqref{FE} is conserved (because the Einstein tensor is divergence free) and the standard matter EMT is also conserved, $T^R_{ab}$ does not have to be \cite{Carloni:2007yv}. In principle, the total EMT in \eqref{FE} can be written as the sum of two conserved quantities:
\begin{equation}
T^{tot}_{ab} = T^m_{ab} + T^{RC}_{ab} \ ,
\end{equation}
with:
\begin{equation}
T^{RC}_{ab} \equiv T^m_{ab} \l( \frac{1}{f_R} -1 \r) + T^R_{ab} \ .
\end{equation}
So, if one insists on writing $\omega_{eff}$ in the forms \eqref{EoSp_density_today} and \eqref{EoSp_density_param}, the total density (Eq. \ref{fRdensity}) can be rewritten as:
\begin{equation}
\rho^{tot} = \rho^m + \rho^{RC} \ ,
\end{equation}
where
\begin{equation}
\rho^{RC} \equiv \rho^m \l( \frac{1}{f_R} -1 \r) + \rho^R \ .
\end{equation}
Needless to say that the corresponding $\omega_{RC}$ and $\Omega_{RC,0}$ have to be determined.


\subsection{The Initial Conditions of Perturbations in $R^n$ Gravity}
\label{IC_Rn}
In the case of $f(R)=R^n$, the evolution of the scale factor, in the presence of a single fluid with standard matter EoS of the form \eqref{EoS} and $\omega = const$, with respect to the cosmic time has been shown to be \cite{Carloni:2004kp,Carloni:2007yv}:
\begin{equation}
a(t) = t^{\frac{2n}{3(1 + \omega)}} \ ,
\label{a_cosmic}
\end{equation}
or, in terms of the conformal time:
\begin{equation}
a(\eta) = \eta^{-\frac{1+\varepsilon}{\varepsilon}} \ ,
\label{a_eta}
\end{equation}
where $\varepsilon \equiv \frac{2n}{3(1 + \omega)} -1$. Using equations \eqref{Ricci-a} and \eqref{a_eta} to substitute for the Ricci scalar and the Hubble parameter in the right hand side of \eqref{sigma_power}, we get a more convenient expression for the parameter $m$:
\begin{equation}
m = 1 + \frac{1-n}{1+\varepsilon} \ ,
\end{equation}
and subsequently for the equation of motion \eqref{uwave}:
\begin{equation}
u_k'' + \l[k^2 + m \l( \frac{\varepsilon + 1}{\varepsilon^2} \r) \l( n-2 \r) \eta^{-2} \r] u_k = 0 \ .
\label{EoM}
\end{equation}
The last equation is equivalent to:
\begin{equation}
u_k'' + \l[k^2 - \frac{(a^{-m})''}{a^{-m}} \r] u_k = 0 \ .
\end{equation}

The initial conditions are set up deep in the radiation dominated era, so $\omega=\frac{1}{3}$. It is worth mentioning that, in this case, equation \eqref{a_eta} has a mathematical singularity at $\varepsilon = 0$ corresponding to $n=2$ \cinc \cinc \cinc \footnote[\countval]{For every $R^n$ model with $\frac{3}{2} \leq n \leq 2$, equation \eqref{a_eta} is ill-posed at some point in the transition from the radiation dominated era to the matter dominated era.}. The $R^2$ term in the gravitational Lagrangian has predictable cosmological consequences as the resulting theory is conformally equivalent to GR plus a scalar field \cite{Capozziello:2007ec,Barrow:2006xb}.

For all other values of $n \neq 2$, substituting for $m$ and $\varepsilon$ corresponding to $\omega = \frac{1}{3}$ in equation \eqref{EoM} leads to:
\begin{equation}
u_k'' + \l[k^2 - 2 \eta^{-2} \r] u_k = 0 \ .
\label{initial_cond}
\end{equation}
This is the same as the result in \cite{Lewis:2000PhD} for tensor perturbations in GR. The fact that the initial conditions of perturbations in $f(R) = R^n$ gravity are the same as in GR is a surprising result at first. But if one remembers the condition \eqref{R_rad} and that we took the curvature parameter $K=0$, the GR initial conditions are actually expected.
 
\section{Numerical Simulations}
\subsection{CAMB}
\label{camb}
We now have the necessary components to begin discussing how to numerically compute the contribution of tensor perturbations to anisotropies in the CMB. For that we use the code for anisotropies in the microwave background (CAMB)  described in \cite{Lewis:2000PhD,camb_notes,camb}.

One way was to use a modified version of CAMB described in \cite{camb_ppf}. It implements a Parametrized Post-Friedmann (PPF) prescription for the dark energy perturbations \cite{Hu:2007pj,Fang:2008sn} allowing it to  take in a time dependent EoS through an input file containing values of $\omega(a)$ vs. $a(t)$. This package provides a way to ``trick'' CAMB into taking background files externally generated for $R^n$ gravity with no cosmological constant and use them as dark energy contribution. However, the curvature fluid does not necessarily evolve in the same way as dark energy and therefore, the implementation is not adequate for our purpose. 

The original CAMB package supports smooth dark energy models with constant equation of state. The idea is to set the density of dark energy to zero and consider a universe containing only dust, radiation, and the curvature fluid. 

\subsection{$R^n$ Modifications to CAMB}
The \fR background is implemented by feeding CAMB a fitting function of $\frac{\text d \eta}{\text d a}$ obtained from the numerical solution of Eq. \eqref{background}. We have proved in subsection \ref{IC_Rn} that the initial conditions are the same as GR for values of $n \neq 2$. Thus CAMB is left unchanged at this level. The remaining part is to modify the tensor perturbation evolution sub-module according to equations \eqref{anis_decomp} and \eqref{anis_Rn_rho} \cinc \footnote[\countval]{see Appendix \ref{camb_modifs}.}. 

Numerical values for $\omega_{eff}(a)$ vs. the scale factor $a(t)$ were obtained from a previous work \cite{Shosho}. A fitting function of $\omega(a) = Polynomial(a)$ allowed easy calculation of $\omega$ and $ \frac{\text d \omega}{\text d \eta} = \frac{\text d \omega}{\text d a} a'$ within CAMB. The goodness of the fit to a quartic polynomial was around $10^{-4}$. However, the fact that there is a background file for each value of $n$ imposed minor, but individual, modifications in the two lines of code concerning $\omega$ and $\text d \omega / \text d \eta$. 

The results of the code executions are presented and discussed in the following chapter.



\chapter{Results and Conclusions} 
\label{discussion}


\ifpdf
    \graphicspath{{6/figures/PNG/}{6/figures/PDF/}{6/figures/}}
\else
    \graphicspath{{6/figures/EPS/}{6/figures/}}
\fi


\section*{CAMB Runs for a GR Background}
\begin{figure}[htbp]
\begin{center}
\includegraphics[width=0.8\textwidth]{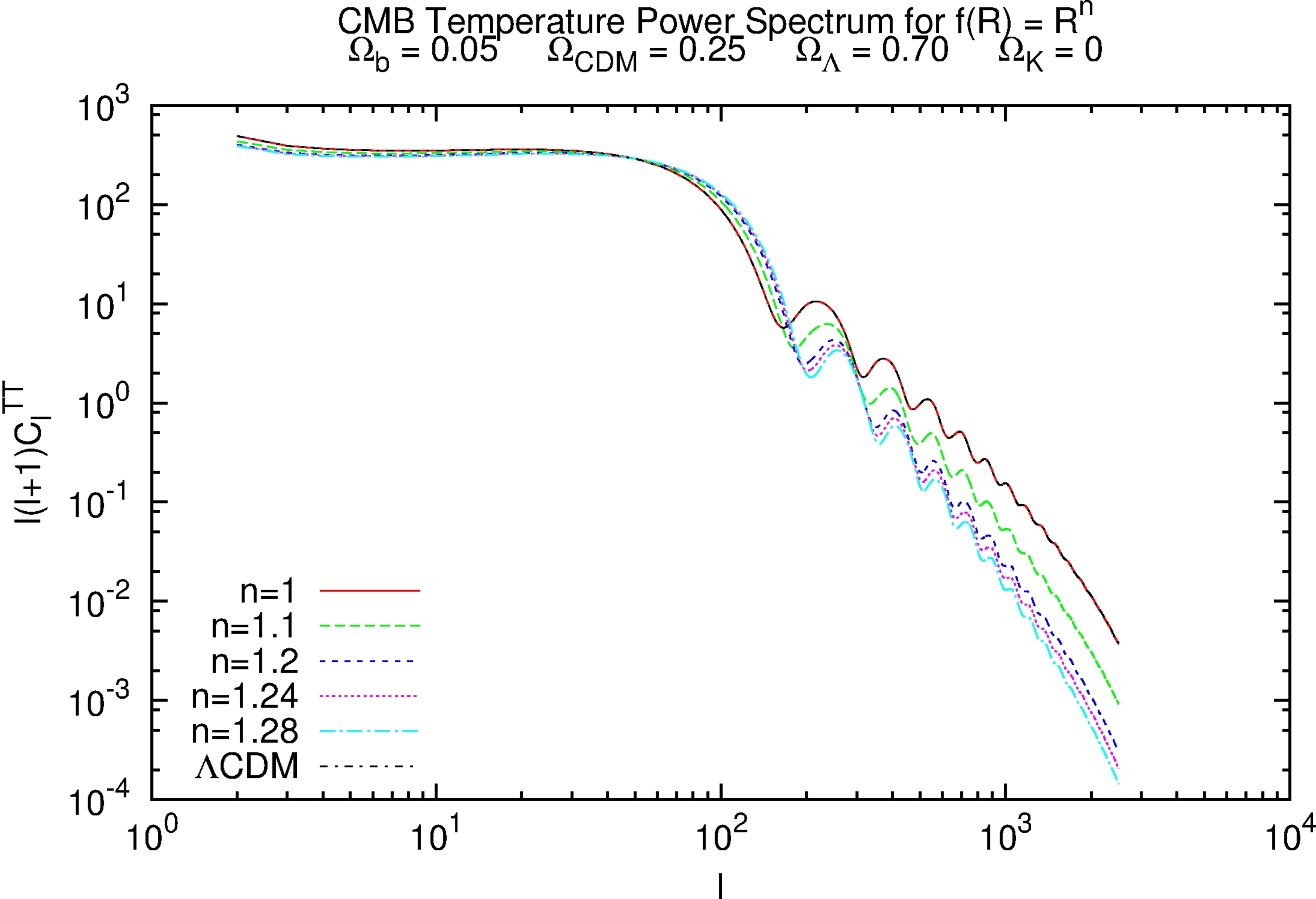}
\caption[Ctt]{\tiny The temperature power spectra for tensor perturbations in $f(R) = R^n$ gravity. The background is GR FLRW. The Hubble constant is chosen to be $H_0=70\, \text{km~s}^{-1} \text{Mpc}^{-1}$. No secondary anisotropies or reionisation effects were considered. The plotted values of $n$ are $1,\ 1.1,\ 1.2,\ 1.24,\ \text{and}\ 1.28$.} 
\label{ctt_gr}
\end{center}
\end{figure}

\begin{figure}[htbp]
\begin{center}
\includegraphics[width=0.8\textwidth]{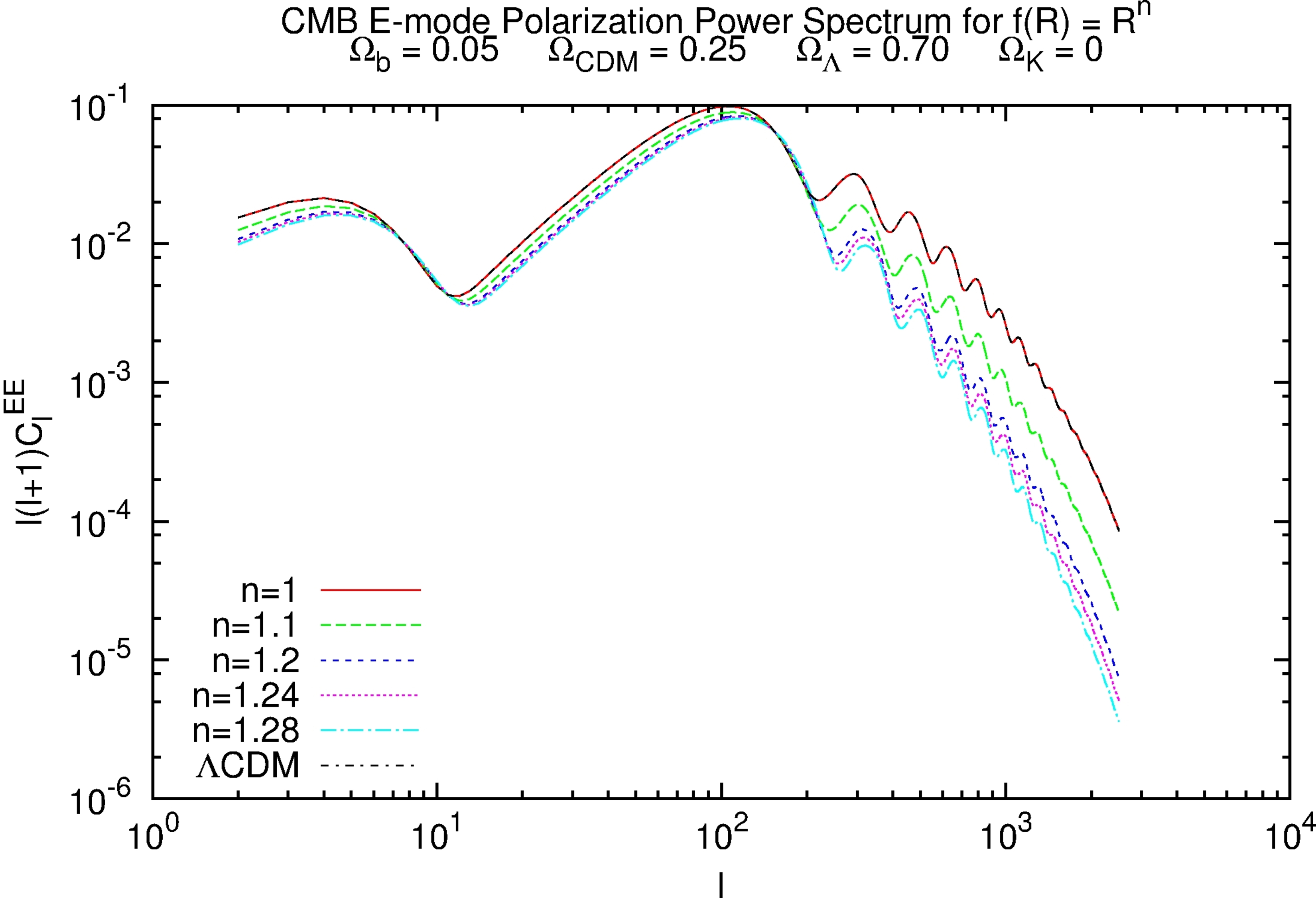} 
\caption[Cee]{\tiny The $E$-mode polarisation power spectra for the same parameters presented in Figure \ref{ctt_gr}.}
\label{cee_gr}
\end{center}
\end{figure}

We start by considering modifications to CAMB where we keep the background the same as general relativity. This allows us to directly compare the influence of the first order modified evolution equations. Figures \ref{ctt_gr} and \ref{cee_gr} respectively show the $TT$ and $EE$ power spectra for $f(R)=R^n$, $n=1,\ 1.1,\ 1.2,\ 1.24,\ \text{and}\ 1.28$. For comparison, the power spectrum for $\Lambda$CDM is also plotted. It was obtained by running CAMB on the same background but with the original GR evolution equations. The density parameters are taken to be $\Omega_b=0.05$, $\Omega_{CDM}=0.25$, and $\Omega_{\Lambda}=0.70$.

A few conclusions can be inferred directly from the power spectra. We start by mentioning that the curve for $\Lambda$CDM and for $R^n = R$ are identical. Also, one notices that the features of the power spectra are shifted more and more towards small scales with increasing power of $R$. The departure from GR increases with increasing power of $R$ as expected from the term $n-1$ in the numerator of the RHS of equation \eqref{curvature_anis_tensor_k_Rn}. It is clear from equation \eqref{curvature_anis_tensor_k} that the curvature anisotropic stress varies with the wavenumber $k$. Therefore, the total perturbations in $f(R)$, unlike GR, are scale dependent. The power difference between the spectra for different values of $n$ becomes larger after multipole $\approx 160$. Finally, the power decreases with increasing values of $n$ except for the interval $60 \lesssim l \lesssim 160$ in the $TT$ power spectra and $140 \lesssim l \lesssim 200$ in the $EE$ power spectra where the opposite happens \cite{Bourhrous:2012kr}.

\section*{The \Rn Background Solutions}
\begin{figure}[htbp]
\begin{center}
\includegraphics[width=0.85\textwidth]{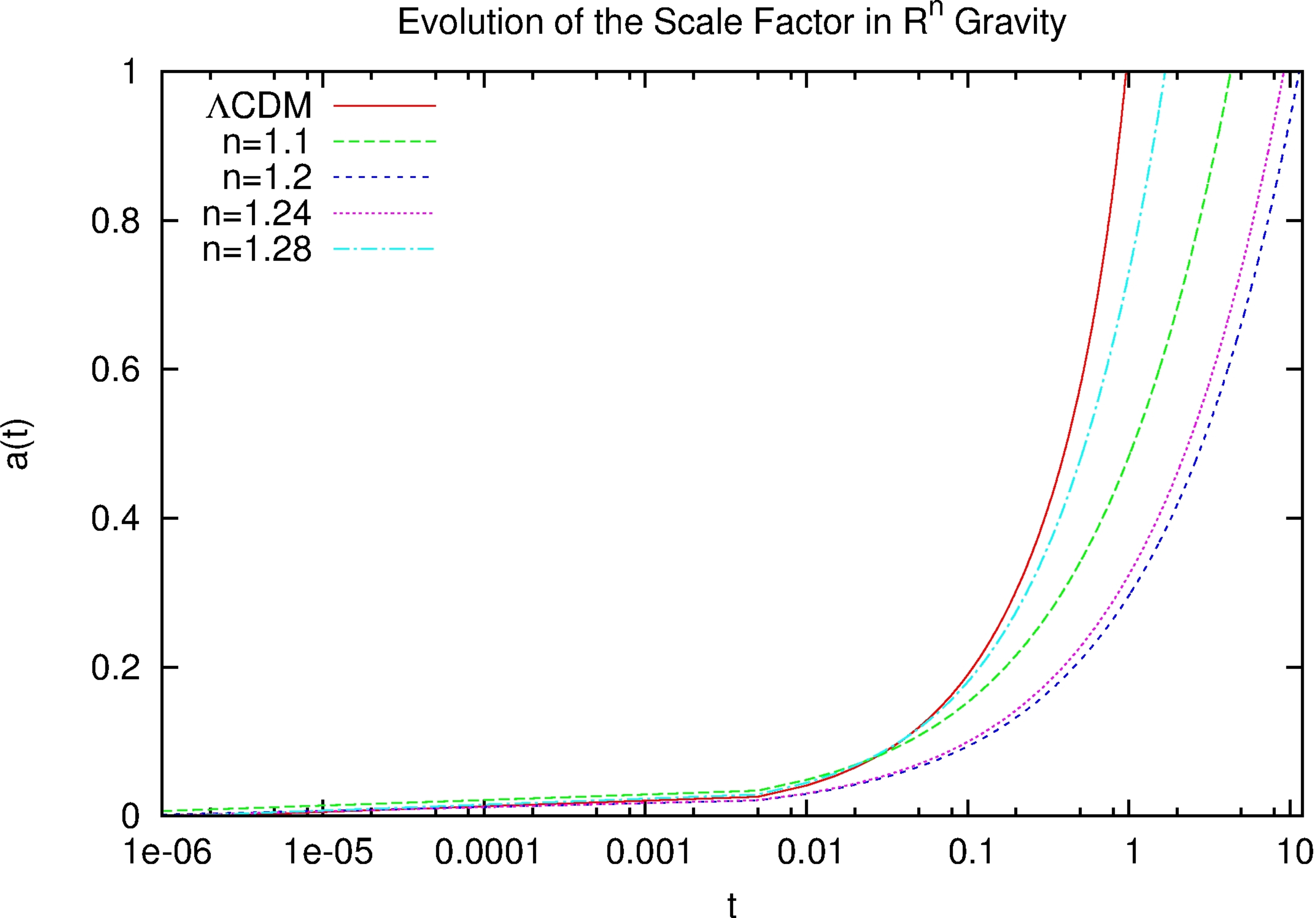}
\caption[Ctt]{\tiny The evolution of the scale factor in $f(R) = R^n$ gravity for $n = \{1,\ 1.1,\ 1.2,\ 1.24,\ 1.28\}$. A cosmological constant contribution was only considered in the $\Lambda$CDM case. The Hubble constant is chosen to be $H_0=70\, \text{km~s}^{-1} \text{Mpc}^{-1}$.} 
\label{at}
\end{center}
\end{figure}

Introducing the corresponding $R^n$ background for each of the considered values of $n$ requires the solutions of equation \eqref{background}. Figure \ref{at} shows the evolution of the scale factor for $n = \{1.1,\ 1.2,\ 1.24,\ 1.28\}$ compared to its evolution in $\Lambda$CDM. 

Assuming that the radiation density is negligible in the present and the universe is filled with dust only, one can see that the cosmic time is not usually unity when the scale factor is unity, as it would expected from to equation \eqref{a_cosmic}. This is not surprising when one recalls that this relation is valid only in the presence of one fluid; but instead there are two, dust and the curvature fluid.


\section*{CAMB Limitations}
Although CAMB is a useful package for GR, it is inadequate for solving fourth, or higher, order differential equations despite the community's considerable efforts to acquaint it to higher order modified theories of gravity. In the context of \fR gravity for example, if the polynomial fitting functions of $\omega(a)$ were not to be used, one would have to deal with up to fifth derivatives of the scale factor; the implementation of which would possibly require massive modifications to the code. 

The most considerable problem is the incompatibility of externally solved background evolution with CAMB. The modification of the background evolution module from GR to \Rn turned out to be difficult. For example, direct assignment of a function of the scale factor to the $(a')^{-1}$ variable does not work.

\section*{Future Work}
The work currently in progress, \cite{abdelwahab:2012}, aims to implement the $R^n$ background properly within CAMB. Power spectra will then be obtained with the adequate background evolution and only then provide a complete picture of CMB tensor anisotropies in \Rn gravity. But the limitations stated in the previous section call for a specialised $R^n$ code. 

In addition to tensor perturbations, scalar perturbations need to be considered as well for a more complete description of CMB anisotropies in \Rn gravity. This work can also be extended further on many other levels such generalising it to non-zero curvatures, including a study of CMB polarisation, and/or considering other \fR models.

\section*{Summary}
Solving for the CMB tensor anisotropies in \fR gravity consisted of three parts. First we derived the initial conditions for tensor perturbations in \fR theories and then in the more special case of $R^n$ gravity. In the latter, the initial conditions were found to be similar to those of GR for all values of $n \neq 2$ which was unexpected since GR and $R^n$ gravity have different backgrounds. Next we derived the equations for the evolution of the tensor mode perturbations in \fR gravity and then more specifically in $R^n$ gravity. In particular, the contribution of the curvature anisotropic stress to the shear was established. Finally, we worked out the equation for the evolution of the scale factor in $R^n$ gravity and obtained numerical solutions. 

In parallel, we modified the evolution equations in CAMB accordingly while keeping the original initial conditions i.e. the ones for GR. The $R^n$ background evolution is yet to be implemented within CAMB and only the GR background has been used so far. Equations \eqref{gwsk}, \eqref{background}, \eqref{curvature_anis_tensor_k}, \eqref{anis_Rn_rho}, and \eqref{initial_cond} together with figures \ref{ctt_gr}, \ref{cee_gr} and \ref{at} are the main results of this work.

\appendix
%
\noappendicestocpagenum
\addappheadtotoc
\chapter{The Propagation Equations}
\label{Propagation_eqs_deriv}

\section{The Ricci Propagation Equations}
Equation \eqref{vector_curv} states a relation between the second derivative of a vector and the Riemann tensor of the underlying geometry.
%
%
Contracted with $u^c$, it becomes:
\begin{align}
\left( u_{a;d} \right)\dot{} - u_{a;cd} u^c &= R_{abcd} u^b u^c \ , \nn\\
\Longleftrightarrow \left( u_{a;d} \right)\dot{} - \dot u_{a;d} + u_{a;c} u^c_{\ph c;d} &= R_{abcd} u^b u^c \ . \nn
\end{align}

Projecting leads to the Ricci propagation equations for the spatial derivative of the velocity vector:
\begin{equation}
h_a{}^c h_b{}^d ( \D_d u_c ) \dot{} - \dot u_a \dot u_b - h_a{}^c h_b{}^d \dot u_{c;d} + \D_c u_a \D_b u^c = R_{amnb} u^m u^n \ .
\label{Ricci_propagation_general}
\end{equation}
Expanding and making use of the Einstein field equations \eqref{EFE}, the trace and trace free decomposition of the Riemann tensor \eqref{Riemann_decomp}, the velocity decomposition equation \eqref{velocity_decomp} and equation \eqref{EMT} for the EMT, one gets to:
\begin{align}
0 & = h_a{}^c h_b{}^d ( \dot \sigma_{cd} + \dot \omega_{cd} ) + \frac{1}{3} h_{ab} \l( \dot \Theta + \frac{1}{3} \Theta^2 \r) + \frac{2}{3} \Theta ( \sigma_{ab} + \omega_{ab} ) + \sigma_{ac} \sigma^c_{\ph cb} + \omega_{ac} \omega^c_{\ph cb} \nn\\
& + \omega_{ac} \sigma^c_{\ph cb} + \sigma_{ac} \omega^c_{\ph cb} - \dot u_a \dot u_b - h_a{}^c h_b{}^d \dot u_{c;d} + E_{ab} - \frac{1}{3} \Lambda h_{ab} + \frac{1}{6} h_{ab} (\rho + 3 p) - \frac{1}{2} \pi_{ab} \ .
\label{Ricci_propagation_expanded}
\end{align}

To get a set of usable formulae, the last equation is decomposed into its trace, symmetric trace free, and antisymmetric parts.

\subsection{The Trace of the Ricci Propagation Equations}
Taking the trace of the previous equation leads to:
\begin{equation}
\dot \Theta  = - \frac{1}{3} \Theta^2 - 2 \left( \sigma^2 - \omega^2 \right) + \Lambda - \frac{1}{2} \left( \rho + 3p \right) + \dot u^a_{\phantom{a};a} \ ,
\label{raychaudhuri}
\end{equation}
where $\sigma^2 \equiv \frac{1}{2} \sigma_{ab} \sigma^{ab}$ and $\omega^2 \equiv \frac{1}{2} \omega_{ab} \omega^{ab}$ are the magnitudes of $\sigma_{ab}$ and $\omega_{ab}$, respectively. This is the Raychaudhuri equation \cite{Raychaudhuri:1953yv}. It describes the evolution of the world lines in each point of spacetime due to its energy content and kinematic setup. Positive standard matter density, non-zero shear, and positive expansion scalar tend to draw the world lines closer (collapse) while a positive cosmological constant, non-zero vorticity, and a positive acceleration divergence pulls the world lines apart (expansion). This becomes more obvious after substituting $\Theta = 3 \frac{\dot a}{a}$ into the Raychaudhuri equation:
$$ 3 \frac{\ddot a}{a} = - 2 \left( \sigma^2 - \omega^2 \right) + \Lambda - \frac{1}{2} \left( \rho + 3p \right) + \dot u^a_{\phantom{a};a} \ .$$

In an FLRW universe, the effect of the shear balances that of the vorticity and the velocity divergence combined:
$$ \dot u^d_{\phantom{d};d} + 2 \omega^2 = 2 \sigma^2 \ , $$
where we have used the acceleration Friedmann equation \eqref{friedmann_ddot} to get the last relation.

The Raychaudhuri equation is a general description of gravitational collapse (or expansion). Take for example the case of a static $(\sigma^2 = \omega^2 = \Theta = \dot \Theta = 0)$ relativistic star. Assuming a vanishing cosmological constant, equation \eqref{raychaudhuri} becomes:
$$ \dot u^a_{\phantom{a};a} = \frac{1}{2} \left( \rho + 3p \right) \ , $$ 
which is indeed the condition for hydrostatic equilibrium. The very pressure keeping the star from collapsing contributes to the forces tending to collapse it \cite{Ellis:2009go}. 

\subsection{The Symmetric Trace Free Ricci Propagation Equations}

If we define a new rank two tensor $t_{ab}$ being the RHS of equation \eqref{Ricci_propagation_expanded}, then $t_{ab} = 0$. The symmetric part of the Ricci propagation equations is isolated via:
$$t_{(ab)} = \frac{1}{2} (t_{ab} + t_{ba}) \ . $$ 
After some calculation, we get:
\begin{align}
t_{(ab)} = 0 = & \ h_a{}^c h_b{}^d \dot \sigma_{cd}  + \frac{1}{3} h_{ab} \l[ \dot u^a_{\phantom{a};a} - 2 \l( \sigma^2 - \omega^2 \r) \r] + \frac{2}{3} \Theta \sigma_{ab} + \sigma_{ac} \sigma^c_{\ph cb} + \omega_{ac} \omega^c_{\ph cb} \nn\\
& - \dot u_a \dot u_b - h_a{}^c h_b{}^d \dot u_{(c;d)} + E_{ab} - \frac{1}{2} \pi_{ab} \ ,
\end{align}
where the Raychaudhuri equation has been used to substitute for the $\dot \Theta + \frac{1}{3} \Theta^2$ term.

The symmetric trace free part can now be obtained by subtracting the trace:
$$t_{\la ab \ra} = t_{(ab)} - \frac{1}{3} h_{ab} t^a_{\ph aa} \ . $$
$t^a_{\ph aa}$ is nothing but the terms in the Raychaudhuri equation \eqref{raychaudhuri}. Then:
\begin{align}
t_{\la ab \ra} = 0 = & \ h_a{}^c h_b{}^d \dot \sigma_{cd} + \frac{2}{3} \Theta \sigma_{ab} + \sigma_{ac} \sigma^c_{\ph cb} + \omega_{ac} \omega^c_{\ph cb} \nn\\
& - \dot u_a \dot u_b - h_a{}^c h_b{}^d \dot u_{(c;d)} + E_{ab} - \frac{1}{2} \pi_{ab} \ .
\end{align}

\subsection{The Anti-Symmetric Trace Free Ricci Propagation Equations}
Similarly, the antisymmetric part is isolated by:
$$t_{[ab]} = \frac{1}{2} (t_{ab} - t_{ba}) \ .$$ 
One gets:
\begin{equation}
t_{[ab]} = 0 = h_a{}^c h_b{}^d \dot \omega_{cd} + \frac{2}{3} \Theta \omega_{ab} + \sigma_{ac} \omega^c_{\ph cb} + \omega_{ac} \sigma^c_{\ph cb} - h_a{}^c h_b{}^d \dot u_{[c;d]}
\end{equation}

\subsubsection*{The Time Evolution of the Vorticity Vector}
%

Multiplying the previous equation by $\varepsilon^{kab}$ gives:
%
%
%
\begin{align}
& \varepsilon^{kcd} \dot \omega_{cd} - \varepsilon^{kcd} \dot u_{[c;d]} + \frac{2}{3} \varepsilon^{kab} \Theta \omega_{ab} +
\varepsilon^{kab} \sigma_{ac} \omega^c_{\phantom{c} b} + \varepsilon^{kab} \sigma^c_{\ph cb} \omega_{ac} = 0 \ ,  \nn\\
\Longleftrightarrow \quad & 2 \dot \omega^k - \varepsilon^{kcd} \dot u_{[c;d]} + \frac{4}{3} \Theta \omega^k + \varepsilon^{kab} \sigma_{ac}
\omega^c_{\phantom{c} b} + \varepsilon^{kab} \sigma^c_{\ph cb} \omega_{ac} = 0 \ , \nn\\
\Longleftrightarrow \quad & \dot \omega^k - \frac{1}{2} \varepsilon^{kcd} \dot u_{[c;d]} + \frac{2}{3} \Theta \omega^k + \varepsilon^{kab} \sigma_{ac} \omega^c_{\phantom{c} b} = 0 \ .
\end{align}

\section{Contracted Second Bianchi Equations for a Perfect Fluid}

The Bianchi identities, $R{ab[cd;e]} = 0$, imply \cite{Ellis:2009go,Bertschinger:1993xt}:
\begin{subequations}
\begin{align}
C^{abdc}_{\phantom{abcd};d} &= R^{c[a;b]} - \frac{1}{6} g^{c[a} R^{;b]} \label{weyl_div_R}\\
\Longleftrightarrow C^{abdc}_{\phantom{abcd};d} &= T^{c[a;b]} - \frac{1}{3} g^{c[a} T^{;b]} \label{weyl_div_T}
\end{align}
\end{subequations}
The move from \eqref{weyl_div_R} the first to \eqref{weyl_div_T} involved the use of the field equations. Substituting \eqref{weyl} into \eqref{weyl_div_T} then performing different time and space projections on different indices leads to the contracted second Bianchi equations. 

\subsection*{The Time Evolution of the Electric Weyl Tensor}
Keeping in mind that $\eta^{abpq} \eta_{cdrs} = 4! \delta^a_{[c} \delta^b_d \delta^p_r \delta^q_{s]}$ \cite{Schutz}, multiplying equation \eqref{weyl_div_T} by $u_a$ and projecting with $h^m_{\ph mb} h^n_{\ph nc}$ leads to, after some long but straight forward algebra:

\begin{eqnarray}
h^m_{\ph mb} h^n_{\ph nc} \dot E^{bc} - h^m_{\ph mb} \epsilon^{nds} H^b_{\ph bs;d} + \Theta E^{mn} + \omega_{rd} h^n_{\ph nc} \eta^{cdrs} H^m_{\ph ms} - 2 \dot u_p u_a H_q^{\ph q(m}\eta^{n)apq} \nonumber \\ + h^{mn} \sigma_{pd} E^{pd} - 3 \sigma_p^{\ph p(m} E^{n)p} - \omega^{(m}_{\phantom{(m}p} E^{n)p} = - \frac{1}{2} \sigma^{mn} \left( \rho + p \right)	\ .
\end{eqnarray}

\subsection*{The Divergence of the Electric Weyl Tensor}
Projecting equation \eqref{weyl_div_T} with $h^t_m h^m_a h_{bc}$ gives:

\begin{equation}
h^t_{\ph tf} h^q_{\ph qe} E^{ef}_{\ph{ef};q} + \dot u_e E^{et} + 2 \omega^s H^t_{\ph ts} - \epsilon^{tbq} \sigma_{bd} H^d_{\ph dq} = \frac{1}{3} h^t_{\ph ta} \rho^{;a}	\ .
\end{equation}

\subsection*{The Time Evolution of the Magnetic Weyl Tensor}
Multiplying equation \eqref{weyl_div_T} by $\frac{1}{2} h^{mp} h^{nc} \epsilon_{pab}$ leads to:
\begin{align}
h^m_{\ph mp} h^n_{\ph nq} \dot H^{pq} -h^{p(m} \epsilon^{n)ab} E_{pb;a} -2 A_q E_a^{\ph a(m} \epsilon^{n)qa} + \Theta H^{mn} + h^{mn} \sigma^{pq} H_{pq} & \nn\\
- 3H^{p(m} \sigma^{n)}_{p} + H^{p( m}\omega^{n)}_{\phantom{n0}p} & = 0	\ .
\end{align}

\subsection*{The Divergence of the Magnetic Weyl Tensor}
Multiplying equation \eqref{weyl_div_T} by $\frac{1}{2} h^{mp} u_c \epsilon_{pab}$ gives:
\begin{equation}
h^m_{\ph mp} h^n_{\ph nq} H^{pq}_{\phantom{pq};n}  -\epsilon^{m p q} \sigma_{pa}E^a_{\ph aq} + 3 E^m_{\ph mn}\omega^n\nonumber   =-(\rho+p)\omega^m \ .
\end{equation}

\chapter{The \fR Field Equations in the Metric Formalism}
\label{FE_derivation}

\section{The $f(R)$ Field Equations}
Considering the gravitational Lagrangian:
\begin{equation}
\mathcal{L} = f(R) = f \qquad \text{and defining} \qquad f_R \equiv \frac{\delta f}{\delta R} \ , \nn
\end{equation}
the action is then:
\begin{equation}
S = \int_\mathcal{R} f \sqrt{-g} d^4x \ . \nn 
\end{equation}

Variation of the gravitational action with respect to the metric gives:
\begin{align*}
\delta S &= \int_\mathcal{R} \delta \left( f \sqrt{-g} \right) d^4x \nn\\
&= \int_\mathcal{R} \left(f \delta \sqrt{-g} + \delta f  \sqrt{-g}\right) d^4x  \nn\\
&= \int_\mathcal{R} \left(-\frac{1}{2} \sqrt{-g} g_{ a  b } \delta g^{ a  b } f + f_R \delta R \sqrt{-g}\right) d^4x \nn\\
&= \int_\mathcal{R} \left[ -\frac{1}{2} f g_{ a  b } \delta g^{ a  b }  + f_R \left( R_{ a  b } + g_{ a  b } \Box - \nabla_ a  \nabla_ b  \right) \delta g^{ a  b } \right] \sqrt{-g} d^4x \qquad (\mathrm{Proof:}\ \S \ \ref{proofdeltaR}) \nn\\
\end{align*}
\begin{align}
&= \int_\mathcal{R} -\frac{1}{2} f g_{ a  b } \delta g^{ a  b }  \sqrt{-g} d^4x \nn\\
& + \int_\mathcal{R} g_{ a  b } \Box \delta g^{ a  b } f_R \sqrt{-g} d^4x \nn\\
& - \int_\mathcal{R} \nabla_ a  \nabla_ b  \delta g^{ a  b } f_R \sqrt{-g} d^4x \nn\\
& + \int_\mathcal{R} R_{ a  b } \delta g^{ a  b } f_R \sqrt{-g} d^4x \nn\\
&= \int_\mathcal{R} -\frac{1}{2} f g_{ a  b } \delta g^{ a  b }  \sqrt{-g} d^4x \nn\\
& + \int_\mathcal{R} g_{ a  b } \delta g^{ a  b } \Box f_R \sqrt{-g} d^4x + \int_\mathcal{R} \nabla^ d  \mathcal{M}_ d  \sqrt{-g} d^4x \qquad (\mathrm{Proof:}\ \S \ \ref{proofsurftermM}) \nn\\
& - \int_\mathcal{R} \delta g^{ a  b } \nabla_ a  \nabla_ b  f_R \sqrt{-g} d^4x - \int_\mathcal{R} \nabla_ d  \mathcal{N}^ d  \sqrt{-g} d^4x  \qquad (\mathrm{Proof:}\ \S \ \ref{proofsurftermN}) \nn\\
& + \int_\mathcal{R} R_{ a  b } \delta g^{ a  b } f_R \sqrt{-g} d^4x \nn\\
&= \int_\mathcal{R} -\frac{1}{2} f g_{ a  b } \delta g^{ a  b }  \sqrt{-g} d^4x \nn\\
& + \int_\mathcal{R} g_{ a  b } \delta g^{ a  b } \Box f_R \sqrt{-g} d^4x + \int_{\partial \mathcal{R}} n^ d   \mathcal{M}_ d  \sqrt{h} d^3y \nn\\
& - \int_\mathcal{R} \delta g^{ a  b } \nabla_ a  \nabla_ b  f_R \sqrt{-g} d^4x - \int_{\partial \mathcal{R}} n_ d   \mathcal{N}^ d  \sqrt{h} d^3y \nn\\
& + \int_\mathcal{R} R_{ a  b } \delta g^{ a  b } f_R \sqrt{-g} d^4x \nn\\
&= \int_\mathcal{R} \left[ -\frac{1}{2} g_{ a  b } f + \left( R_{ a  b } + g_{ a  b } \Box - \nabla_ a  \nabla_ b  \right) f_R \right] \delta g^{ a  b } \sqrt{-g} d^4x \nn\\
&+ \int_{\partial \mathcal{R}} n^ d   \mathcal{M}_ d  \sqrt{h} d^3y - \int_{\partial \mathcal{R}} n_ d   \mathcal{N}^ d  \sqrt{h} d^3y \ . \nn\\
\end{align}

Adding the matter term and assuming that the surface term vanishes, $\delta S$ becomes:
$$\delta S = \frac{1}{2} \int_\mathcal{R} \left[ -\frac{1}{2} g_{ a  b } f + \left( R_{ a  b } + g_{ a  b } \Box - \nabla_ a  \nabla_ b  \right) f_R \right] \delta g^{ a  b } \sqrt{-g} d^4x + \int_\mathcal{R} \delta \mathcal{L}_m \sqrt{-g} d^4x \ , $$  
where $\mathcal{L}_m$ is the Lagrangian associated with the matter field:
$$\delta \mathcal{L}_m = - \frac{\sqrt{-g}}{2}\delta g^{ a  b } T_{ a  b } \ ,$$
with $T_{ a  b }$ being the energy-momentum tensor.

Requiring the stationary action condition $\delta S = 0$, we get:
$$\int_\mathcal{R} \left[ -\frac{1}{2} g_{ a  b } f + \left( R_{ a  b } + g_{ a  b } \Box - \nabla_ a  \nabla_ b  \right) f_R -  T_{ a  b } \right] \delta g^{ a  b } \sqrt{-g}  d^4x = 0 \ , $$ 
which means that, for an arbitrary variation $\delta g^{ a  b }$, the integrand vanishes:
$$-\frac{1}{2} g_{ a  b } f + \left( R_{ a  b } + g_{ a  b } \Box - \nabla_ a  \nabla_ b  \right) f_R -  T_{ a  b } = 0 \ , $$
which is equivalent to:
$$f_R R_{ a  b } -\frac{1}{2} g_{ a  b } f - \left(\nabla_ a  \nabla_ b  - g_{ a  b } \Box \right) f_R =  T_{ a  b } \ . $$

After a simple rearrangement of the terms, one gets the field equations in the metric formalism of \fR gravity:
\begin{equation}
G_{ a  b }^{(eff)} =  T_{ a  b }^{(eff)} \ ,
\end{equation}
where
$$G_{ a  b }^{(eff)} \equiv f_R G_{ a  b } = f_R \l( R_{ a  b } -\frac{1}{2} g_{ a  b } R \r) \ , $$
and
$$T_{ a  b }^{(eff)} \equiv T_{ a  b } + \left[ \frac{1}{2} g_{ a  b } f + \left(\nabla_ a  \nabla_ b  - g_{ a  b } \Box \right) f_R  - \frac{1}{2} f_R g_{ a  b } R \right] \ . $$


\vspace{0.5cm}

\noindent \cite{Hobson,Sotiriou:2008rp,Sotiriou:2007yd,Guarnizo:2010xr}.

\newpage
\section{Proofs for Intermediate Steps}
\subsection{$\delta R = R_{ a  b } \delta g^{ a  b } + g_{ a  b } \Box \delta g^{ a  b } - \nabla_ a  \nabla_ b  \delta g^{ a  b }$}
\label{proofdeltaR}

We start from the definition of the Riemann tensor:
$$R^ c _{\phantom{ c } a  e  b } \equiv \partial_ e  \Gamma^ c _{\phantom{ c } a  b } - \partial_ b  \Gamma^ c _{\phantom{ c } a  e }  + \Gamma^ f _{\phantom{ f } a  b } \Gamma^ c _{\phantom{ c } f  e } - \Gamma^ f _{\phantom{ f } a  e } \Gamma^ c _{\phantom{ c } f  b } \ . $$
In a local inertial frame, the affine connection vanishes. The previous equation becomes:
$$R^ c _{\phantom{ c } a  e  b } = \partial_ e  \Gamma^ c _{\phantom{ c } a  b } -  \partial_ b   \Gamma^ c _{\phantom{ c } a  e } \ . $$
For a variation in the connection, $\Gamma^ c _{\phantom{ c } a  b } \rightarrow \Gamma^ c _{\phantom{ c } a  b } + \delta \Gamma^ c _{\phantom{ c } a  b }$, the variation of the Riemann tensor is:
$$\delta R^ c _{\phantom{ c } a  e  b } = \partial_ e  (\delta \Gamma^ c _{\phantom{ c } a  b }) - \partial_ b  (\delta \Gamma^ c _{\phantom{ c } a  e }) \ . $$
But since $\delta \Gamma^ c _{\phantom{ c } a  b }$ is a tensor, the partial derivative can generalised into the covariant derivative:
$$\delta R^ c _{\phantom{ c } a  e  b } = \nabla_ e  (\delta \Gamma^ c _{\phantom{ c } a  b }) - \nabla_ b  (\delta \Gamma^ c _{\phantom{ c } a  e }) \ . $$
This is the Palatini equation. Contracting on $ c $ and $ e $ gives:
$$\delta R_{ a  b } = \nabla_ c (\delta \Gamma^ c _{\phantom{ c } a  b }) - \nabla_ b  (\delta \Gamma^ c _{\phantom{ c } a  c })$$

Now:
\begin{align*}
\delta R &= \delta (g^{ a  b }R_{ a  b }) \ , \\
&= R_{ a  b }\delta g^{ a  b } + g^{ a  b } \delta R_{ a  b } \ , \\
&= R_{ a  b }\delta g^{ a  b } + g^{ a  b } \left[\nabla_ c (\delta \Gamma^ c _{\phantom{ c } a  b }) - \nabla_ b  (\delta \Gamma^ c _{\phantom{ c } a  c })\right]  \ , \\
&= R_{ a  b }\delta g^{ a  b } + \nabla_ c  \left[ g^{ a  b } \delta \Gamma^ c _{\phantom{ c } a  b } -  g^{ a  c } \delta \Gamma^ e _{\phantom{ e } a  e } \right] \ , \\
&= R_{ a  b }\delta g^{ a  b } + \nabla_ c  \left[ g_{ p  q } \nabla^ c  \left( \delta g^{ p  q } \right)- \nabla_ d  \left( \delta g^{ d  c } \right) \right] \ ,  \qquad (\mathrm{Proof:} \ \S \ \ref{proofgdeltaGamma}) \\
&= R_{ a  b } \delta g^{ a  b } + g_{ a  b } \Box \delta g^{ a  b } - \nabla_ a  \nabla_ b  \delta g^{ a  b } \ . \ \  \qquad \qquad \qquad \text{QED}
\end{align*}

\subsubsection{ $g^{ a  b } \delta \Gamma^ c _{\protect \phantom{ c } a  b } - g^{ a  c } \delta \Gamma^ e _{\protect \phantom{ e } a  e } =  g_{ p  q } \nabla^ c  \left( \delta g^{ p  q } \right)- \nabla_ d  \left( \delta g^{ d  c } \right)$}
\label{proofgdeltaGamma}
We have:
\begin{align*}
\delta \Gamma^ c _{\phantom{ c } a  b }  =& \delta \left[ \frac{1}{2} g^{ c  e } (g_{ a  e , b } + g_{ b  e , a } - g_{ a  b , e }) \right] \ , \\
=& \frac{1}{2}\delta g^{ c  e }  (g_{ a  e , b } + g_{ b  e , a } - g_{ a  b , e })+ \frac{1}{2} g^{ c  e }  \left[(\delta g_{ a  e })_{, b } + (\delta g_{ b  e })_{, a } - (\delta g_{ a  b })_{, e }\right] \ , \\
=& \frac{1}{2}\delta g^{ c  e }  (g_{ a  e , b } + g_{ b  e , a } - g_{ a  b , e }) \\
&+ \frac{1}{2} g^{ c  e } \Big[\nabla_ a  \delta g_{ e  b }  + \Gamma^ f _{\phantom{ f } a  e } \delta g_{ f  b } + \Gamma^ f _{\phantom{ f } a  b } \delta g_{ f  e } \\
&+ \nabla_ b  \delta g_{ a  e }  + \Gamma^ f _{\phantom{ f } a  b } \delta g_{ f  e } + \Gamma^ f _{\phantom{ f } b  e } \delta g_{ f  a } \\
&- \nabla_ e  \delta g_{ a  b }  - \Gamma^ f _{\phantom{ f } a  e } \delta g_{ f  b } - \Gamma^ f _{\phantom{ f } e  b } \delta g_{ f  a }\Big] \ , \\
=& \frac{1}{2}\delta g^{ c  e }  (g_{ a  e , b } + g_{ b  e , a } - g_{ a  b , e }) 
+ \frac{1}{2} g^{ c  e } \Big[\nabla_ a  \delta g_{ e  b }  + \nabla_ b  \delta g_{ a  e } + \nabla_ e  \delta g_{ a  b } \Big] \\
&+ g^{ c  e } \delta g_{ f  e } \Gamma^ f _{\phantom{ f } a  b } \ , \\
=& \frac{1}{2}\delta g^{ c  e }  (g_{ a  e , b } + g_{ b  e , a } - g_{ a  b , e }) 
+ \frac{1}{2} g^{ c  e } \Big[\nabla_ a  \delta g_{ e  b }  + \nabla_ b  \delta g_{ a  e } - \nabla_ e  \delta g_{ a  b } \Big] \\
& - g^{ c  e } g_{ f  p } g_{ e  q } \delta g^{ p  q } \Gamma^ f _{\phantom{ f } a  b } \ , \\
=& \frac{1}{2} g^{ c  e } \Big[\nabla_ a  \delta g_{ e  b }  + \nabla_ b  \delta g_{ a  e } - \nabla_ e  \delta g_{ a  b } \Big] \ , \\
=& \frac{1}{2} g^{ c  e } \left[\nabla_ a  \left( - g_{ p  b } g_{ e  q } \delta g^{ p  q } \right) 
+ \nabla_ b   \left( - g_{ p  e } g_{ q  a } \delta g^{ p  q } \right) 
- \nabla_ e  \left( - g_{ p  a } g_{ q  b } \delta g^{ p  q } \right) \right] \ , \\
=& - \frac{1}{2} \left[ g_{ p  b } \delta^ c _ q  \nabla_ a  \left( \delta g^{ p  q } \right) 
+ \delta^ c _ p  g_{ q  a } \nabla_ b   \left( \delta g^{ p  q } \right) 
- g_{ p  a } g_{ q  b } \nabla^ c  \left( \delta g^{ p  q } \right) \right] \ , \\
=& - \frac{1}{2} \left[ g_{ p  b } \nabla_ a  \left( \delta g^{ p  c } \right) 
+ g_{ q  a } \nabla_ b   \left( \delta g^{ c  q } \right) 
- g_{ p  a } g_{ q  b } \nabla^ c  \left( \delta g^{ p  q } \right) \right] \ .
\end{align*}


Contracting on $ c $ and $ b $ gives:
$$\delta \Gamma^ e _{\phantom{ e } a  e } = - \frac{1}{2} g_{ p  q } \nabla_ a  \left( \delta g^{ p  q } \right) \ .$$
Then:
\begin{align*}
g^{ a  b } \delta \Gamma^ c _{\phantom{ c } a  b } - g^{ a  c } \delta \Gamma^ e _{\phantom{ e } a  e } =&
- \frac{1}{2} \Big[ g^{ a  b } g_{ p  b } \nabla_ a  \left( \delta g^{ p  c } \right) 
+ g^{ a  b } g_{ q  a } \nabla_ b   \left( \delta g^{ c  q } \right) \\
&- g^{ a  b } g_{ p  a } g_{ q  b } \nabla^ c  \left( \delta g^{ p  q } \right) - g^{ a  c } g_{ p  q } \nabla_ a  \left( \delta g^{ p  q } \right) \Big] \ , \\
=& g_{ p  q } \nabla^ c  \left( \delta g^{ p  q } \right)- \nabla_ d  \left( \delta g^{ d  c } \right) \ . \qquad \qquad \text{QED}
\end{align*}

\subsection{$ g_{ a  b } f_R \Box \delta g^{ a  b } = g_{ a  b } \delta g^{ a  b } \Box f_R + \nabla^ c  \mathcal{M}_ c $}
\label{proofsurftermM}
\begin{align*}
f_R \Box \delta g^{ a  b } &= f_R \nabla^ c  \nabla_ c  \delta g^{ a  b } \ , \\
&= \nabla^ c  \left[ f_R \nabla_ c  \delta g^{ a  b } \right] - \nabla^ c  f_R \nabla_ c  \delta g^{ a  b } \ , \\
&= \nabla^ c  \left[ f_R \nabla_ c  \delta g^{ a  b } \right] - \nabla_ c  f_R \nabla^ c  \delta g^{ a  b } \ , \\
&= \nabla^ c  \left[ f_R \nabla_ c  \delta g^{ a  b } \right] - \nabla^ c  \left[ \delta g^{ a  b } \nabla_ c  f_R \right] + \delta g^{ a  b } \nabla^ c  \nabla_ c  f_R \ , \\
&= \delta g^{ a  b } \Box f_R + \nabla^ c  \left[ f_R \nabla_ c  \delta g^{ a  b } - \delta g^{ a  b } \nabla_ c  f_R \right] \ . 
\end{align*}
Then
$$g_{ a  b } f_R \Box \delta g^{ a  b } = g_{ a  b } \delta g^{ a  b } \Box f_R + \nabla^ c  \mathcal{M}_ c  \ ,$$
where $\mathcal{M}_ c  \equiv g_{ a  b } f_R \nabla_ c  \delta g^{ a  b } - g_{ a  b } \delta g^{ a  b } \nabla_ c  f_R \ .$  \hspace{1cm} QED

\subsection{$f_R \nabla_ a  \nabla_ b  \delta g^{ a  b } = \delta g^{ a  b } \nabla_ a  \nabla_ b  f_R + \nabla_ a  \mathcal{N}^ a $}
\label{proofsurftermN}
\begin{align*}
f_R \nabla_ a  \nabla_ b  \delta g^{ a  b } 
&= \nabla_ a  \left[ f_R \nabla_ b  \delta g^{ a  b } \right] - \nabla_ a  f_R \nabla_ b  \delta g^{ a  b } \ , \\
&= \nabla_ a  \left[ f_R \nabla_ b  \delta g^{ a  b } \right] - \nabla_ b  \left[ \delta g^{ a  b } \nabla_ a  f_R \right] + \delta g^{ a  b } \nabla_ a  \nabla_ b  f_R \ , \\
&= \delta g^{ a  b } \nabla_ a  \nabla_ b  f_R + \nabla_ a  \left[ f_R \nabla_ b  \delta g^{ a  b } - \delta g^{ a  b } \nabla_ b  f_R \right] \ , \\
&= \delta g^{ a  b } \nabla_ a  \nabla_ b  f_R + \nabla_ a  \mathcal{N}^ a  \ ,
\end{align*}
where $\mathcal{N}^ a  \equiv f_R \nabla_ b  \delta g^{ a  b } - \delta g^{ a  b } \nabla_ b  f_R \ .$ \hspace{1cm} QED
\chapter{$R^n$ Modifications to CAMB}
\label{camb_modifs}

Here, I will talk mainly about the changes made to the equations.f90 CAMB fortran file. Other modifications were made to adjust CAMB into taking new parameters such as the power of the Ricci scalar $n$.

\section{Background}
CAMB calculates the background by integrating the contribution of curvature, dust, radiation, massless neutrinos, and if applicable, $\Lambda$ contribution and massive neutrinos:

\begin{verbatim}
         grhoa2=grhok*a2+(grhoc+grhob)*a+grhog+grhornomass
         if (w_lam == -1._dl) then
           grhoa2=grhoa2+grhov*a2**2
         else
           grhoa2=grhoa2+grhov*a**(1-3*w_lam)
         end if
        if (CP%Num_Nu_massive /= 0) then
!Get massive neutrino density relative to massless
           do nu_i = 1, CP%nu_mass_eigenstates
            call Nu_rho(a*nu_masses(nu_i),rhonu)
            grhoa2=grhoa2+rhonu*grhormass(nu_i)
           end do
        end if

        dtauda=sqrt(3/grhoa2)
\end{verbatim}
then return $dtauda \equiv 1/a'$. 

\section{Initial Conditions}
In \ref{IC_Rn}, we have proved that the initial conditions are the same as for GR. So, we left this part unchanged. 

\section{Evolution}
In the tensor perturbations routine, we added the \Rn contribution to the anisotropic stress according to equations \eqref{anis_decomp} and \eqref{anis_Rn_rho}. Right after the part of the code where CAMB calculates the anisotropic stress for the case of GR, the following lines were added \cinc \footnote[\countval]{The assignments for ``omegaeff'' and ``omegaeffdot'' are the fitting functions for $n=1.28$.}:

\begin{verbatim}
1.    	omegaeff = 0.0415379 - 0.709044*a + 0.880719*a2
                 - 0.599369*a*a2 + 0.173093*a2*a2
2.    	omegaeffdot = (- 0.709044 + 2.0*0.880719*a - 3.0*0.599369*a2 
                     + 4.0*0.173093*a*a2)/dtauda(a)
3.    	Ricci_Scalar = 	(1-3*omegaeff)*8*pi
4.    	Ricci_Scalar_dot = -3*(adotoa*(1+omegaeff)*(1-3*omegaeff) 
                         - omegaeffdot)*8*pi 
5.    	fR_curvature_anisotropy = 
	             shear*(1-power_of_R)*k*Ricci_Scalar_dot/(Ricci_Scalar) 
6.    	rhopi = rhopi/(power_of_R*Ricci_Scalar**(power_of_R-1)) 
               + fR_curvature_anisotropy
\end{verbatim}

The first and second lines introduce the fitting function of $\omega(a) = Polynomial(a)$ and its derivative. The third and fourth lines calculate $R$ and $R'$. The fifth calculates $\pi^R_k$, and finally, the last line is nothing but the fortran implementation of Eq. \eqref{curvature_anis_tensor_k_Rn} or equivalently Eq. \eqref{anis_Rn_rho}. The $\rho a^2$ term in the denominator of these equations is absorbed into ``$rhopi$'' in the code.







\bibliographystyle{bst/utphys} 
\renewcommand{\bibname}{References} 

\bibliography{9_backmatter/references} 




\bibliographystyle{plain} 

\end{document}